\documentclass[a4paper,fleqn]{aa}
\usepackage{inputenc}
\usepackage{amsmath, amssymb}
\usepackage{graphicx}
\usepackage{subcaption}
\usepackage{multicol}
\usepackage{multirow}
\usepackage{booktabs}
\usepackage{natbib}
\usepackage{hyperref}

\newcounter{refcount}

\newcommand{\nmock}{\ensuremath{n_{\textrm{mock}}}}
\renewcommand{\nsim}{\ensuremath{n_{\textrm{sim}}}}
\newcommand{\HRD}{Hertzsprung–Russell diagram}

\newcommand{\feh}{[\rm{Fe/H}]}

\newcommand{\teff}{\ensuremath{T_{\rm{eff}}}}

\newcommand{\myimage}[1]{\begin{center}\includegraphics[angle=0, width=0.9\linewidth]{images/#1}\end{center}}

\newcommand{\myimageTwo}[2]{\begin{center}\includegraphics[angle=0, width=0.45\linewidth]{images/#1}
    \includegraphics[angle=0, width=0.45\linewidth]{images/#2}\end{center}}
\newcommand{\myimageFour}[4]{\begin{center}\includegraphics[angle=0, width=0.45\linewidth]{images/#1}
    \includegraphics[angle=0, width=0.45\linewidth]{images/#2}\end{center}
\begin{center}\includegraphics[angle=0, width=0.45\linewidth]{images/#3}
    \includegraphics[angle=0, width=0.45\linewidth]{images/#4}\end{center}}

\addto\extrasenglish{%
\let\subsectionautorefname\sectionautorefname

}
\title{Ensemble age inversions for large spectroscopic surveys}
\titlerunning{Age inversions.}
\author{Alexey Mints\inst{\ref{inst3}, \ref{inst1}, \ref{inst2}}\thanks{email: amints@aip.de} \and Saskia Hekker\inst{\ref{inst1}, \ref{inst2}} \and Ivan Minchev\inst{\ref{inst3}}}
\authorrunning{A. Mints, S. Hekker and I. Minchev}
\institute{
Leibniz-Institut für Astrophysik Potsdam (AIP), An der Sternwarte 16, 14482 Potsdam, Germany \label{inst3} \and
Max Planck Institute for Solar System Research, Justus-von-Liebig-Weg 3, 37077 Göttingen, Germany \label{inst1} \and
Stellar Astrophysics Centre, Department of Physics and Astronomy, Aarhus University, Ny Munkegade 120, DK-8000 Aarhus C, Denmark \label{inst2}}

\abstract {Galactic astrophysics is now in the process of building a multi-dimensional map of the Galaxy. For such a map, stellar ages are the essential ingredient. Ages are however measured only indirectly by comparing observational 
data with models. It is often difficult to provide a single age value for a given star, as several non-overlapping solutions are possible.} 
{We aim at recovering the underlying log(age) distribution from the measured log(age) probability density function for an arbitrary set of stars.} 
{We build an age inversion method, namely, we represent the measured log(age) probability density function as a 
weighted sum of probability density functions of mono-age populations. Weights in that sum give the underlying log(age) distribution. Mono-age populations are simulated so that the distribution of stars on the $\log g$-\feh\,plane is close to that of the observed sample.} 
{We tested the age inversion method on simulated data, demonstrating that it is capable of properly recovering the true log(age) distribution for a large ($N > 10^3$) sample of stars. The method was further applied to large public spectroscopic surveys. For RAVE-on, LAMOST and APOGEE we also applied age inversion to mono-metallicity samples, successfully recovering age-metallicity trends present in higher-precision APOGEE data and chemical evolution models.}
{We conclude that applying an age inversion method as presented in this work is necessary to recover the underlying age distribution of a large ($N > 10^3$) set of stars. These age distributions can be used to explore for instance age-metallicity relations.}


\date{XXX/YYY}

\begin{document}

\renewcommand{\figureautorefname}{Fig.} 
\renewcommand{\sectionautorefname}{Section} 
\renewcommand{\subsectionautorefname}{Section} 

\maketitle

\section{Introduction}
Stellar ages and distances are important ingredients to compose a reliable picture of the Galaxy. In combination with chemical abundances and kinematics they help us recover the history of star formation and satellite accretion for the Galaxy from its formation to present day.
Large astrometric surveys, of which Gaia \citep{2001A&A...369..339P} with its over $10^9$ objects is now the dominating one, provide us with kinematic information, such as positions, proper motions and parallaxes. To obtain radial velocities and physical properties of the stars, such as temperatures, surface gravities and chemical compositions, spectroscopic data can be used. Modern spectroscopic surveys provide such data for millions of stars.

While distances, chemical abundances and kinematics are typically contained in survey results, ages are not. This is because they are not related directly to any observable parameter and therefore estimating ages is more complicated and in most cases involves models of stellar evolution. \cite{2010ARA&A..48..581S} gives a large overview of different methods used to derive stellar ages. 
One of the methods listed there, the so-called ``isochrone matching'', is useful to derive both ages and distances, and is widely applied to spectroscopic data. 
This method is based on comparison of quantities measured spectroscopically (like the effective temperature \teff, surface gravity $\log g$ and metallicity \feh) to a set of models. A subset of models with parameters close to the observed ones gives an estimate of the age and luminosity of the star. The luminosity combined with visible magnitudes from photometric surveys and extinction values gives an estimate of the distance, independent of the astrometric parallax. 

The recent work of \citet{2019MNRAS.487.3946M} argued that the lack of age information in Galactic Archaeology can lead to severe  misinterpretation of the Milky Way formation and evolution. The authors showed that a number of chemo-kinematical relations used to study the Milky Way are plagued by a phenomenon known as Yule-Simpson’s paradox, which has the effect of erasing or completely reversing the trends seen in mono-age populations when age (or birth radius) is marginalized over.

\begin{figure}
    \myimage{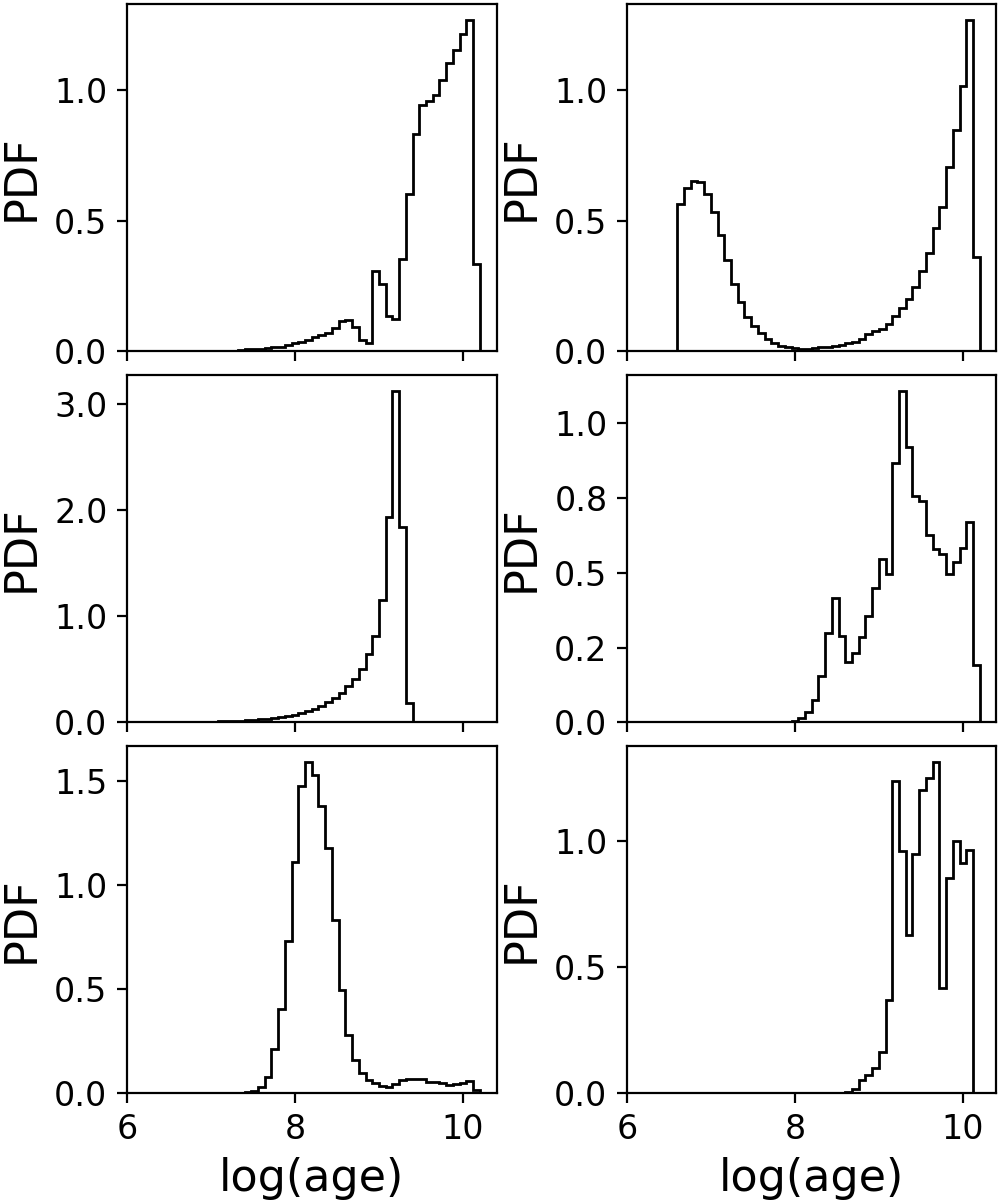}
\caption{Examples of log(age) PDF for various stars. Note that PDFs often show non-Gaussian features: heavy tails, truncation, multiple modes. }\label{fig:age_pdf_examples}
\end{figure}

In \citet[hereafter Paper 1]{UniDAM1} we introduced the implementation of this approach named Unified tool for Distance, Age and Mass estimations (UniDAM\footnote{\url{http://www2.mps.mpg.de/homes/mints/unidam.html}}). This tool uses PARSEC models \citep{PARSEC} and infrared photometry from 2MASS \citep{2006AJ....131.1163S} and AllWISE \citep{2014yCat.2328....0C} surveys to produce probability density functions (PDFs) in distance, log(age) and mass for a given star. A further extension of UniDAM that includes the use of Gaia parallaxes was presented in \citet{UniDAM2} and \citet{UniDAM2a}. The output of UniDAM contains for a given star one or several solutions, with each solution having a unimodal probability density functions (PDF) in log(age), mass and distance (labelled as unimodal sub-PDF, or USPDF). We report for each solution, along with mean, median and mode values, the standard deviation and confidence intervals, also a label indicating the type of best-fitting unimodal function used to fit the USPDF and parameters of that function. This is done in order to overcome problems that are inherent to the isochrone matching method: non-Gaussianity and multi-modality of the produced PDFs.  There are cases when age PDFs are close to being Gaussian, and thus ages can be inferred with high precision. This occurs when high quality data are used and a specific subset of stars is analysed, like it was done, for example, in \citet{2016A&A...590A..32T} with high-resolution spectra of solar twins and in \citet{2017RAA....17....5W} with spectroscopic and asteroseismic data for main sequence turn-off stars. 
In a general case like a large spectroscopic survey, derivation of an age value from an age PDF is often more problematic. This is illustrated in \autoref{fig:age_pdf_examples}, where we show examples of log(age) PDFs for several stars. Some log(age) PDFs are in fact close to exponential functions, which corresponds to flat PDFs in linear ages. For such stars we can put little or no constraints on their age.
An important consequence of these problems is the fact that it is difficult or even impossible to provide for a given star a single age estimate that will be both accurate and precise. 
In the literature mean values \citep[see, for example][]{2016ApJ...817...40F, 2018MNRAS.tmp..326Q} or modes \citep{Xiang_2017} of PDFs were used as age proxies. The implicit assumption is that these proxies are unbiased -- at least in a statistical sense. In this work, we argue that this assumption is not always valid, and we provide a way of reconstructing the age distribution of a stellar population.

\section{PDF quality}\label{sec:pdf}
\begin{figure*}
    \myimage{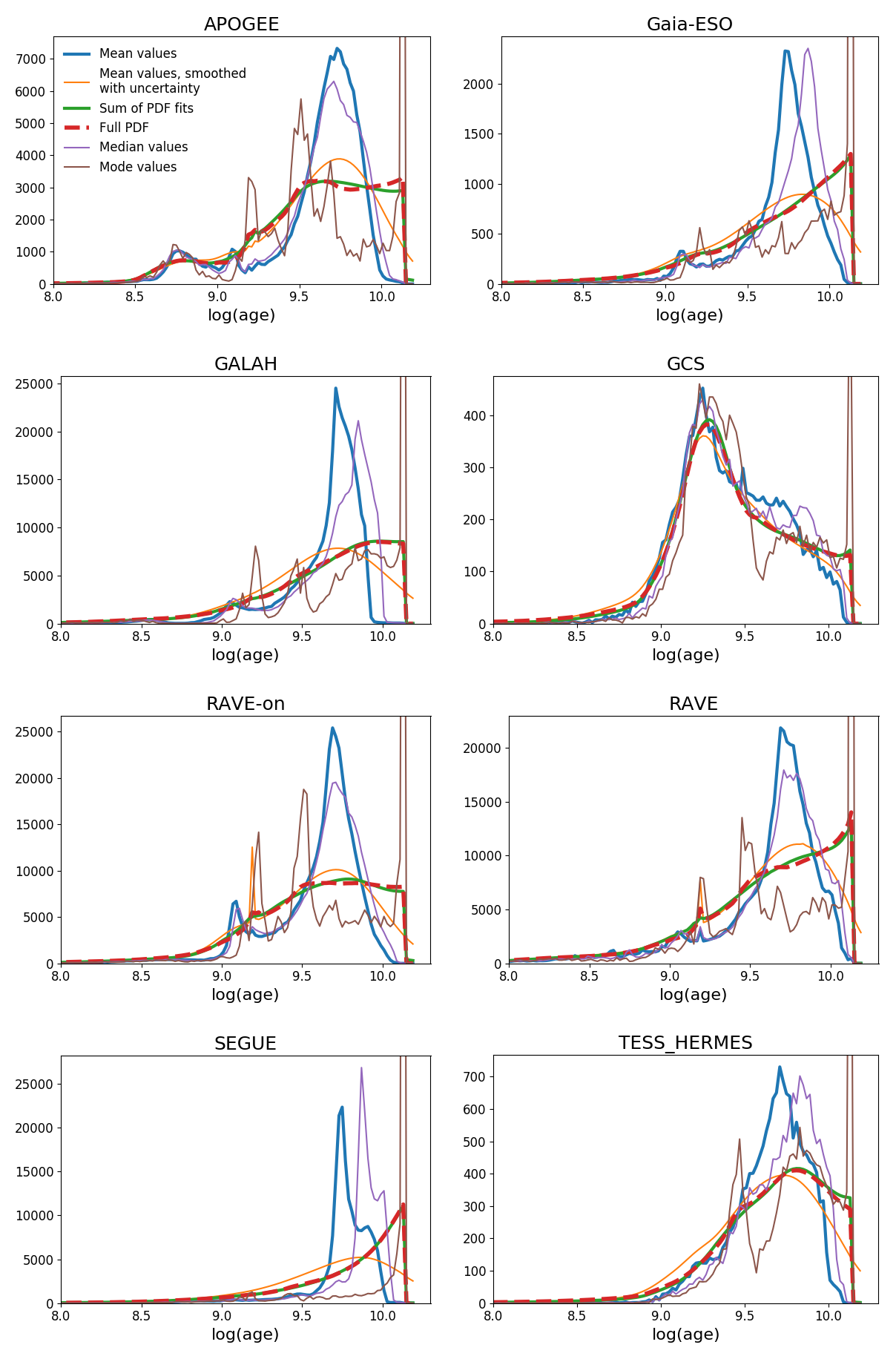}
    \caption{Different representations of log(age) distribution for public spectroscopic surveys. See \autoref{sec:pdf} for a discussion.}\label{fig:surveys}
\end{figure*}

For our task we will be using UniDAM results presented in \citet{UniDAM1}. Fits to log(age) PDFs produced by UniDAM allow us to quickly reconstruct these PDFs for any subset of the survey. Here, we first want to show that these fits give a reliable representation of log(age) PDFs. 
We do that by comparing for each survey the stacked PDF for all stars with the sum of PDF fits produced by UniDAM.
In this section, 
we analyse log(age) distributions to show that using USPDF fits from our catalogue have an advantage over using mean values alone or mean values and uncertainties. This is illustrated in \autoref{fig:surveys}, which shows several representations of log(age) distribution, for every survey processed by UniDAM: 
\begin{itemize}
	\item Histogram of mean (blue line in \autoref{fig:surveys}), median (purple line) or mode (brown line) values of USPDFs for all stars:
	\begin{equation}
	N(\tau_b) = \sum_{i: \tau_i \in (\tau_b - \Delta, \tau_b + \Delta)} w_i,
	\end{equation}
	where $\tau_b$ are centres of bins in log(age), $\Delta$ is the bin half-width
	and $w_i$ is the weight of the USPDF $i$\footnote{In UniDAM, USPDF weight is the fraction of the total PDF contained in the USPDF.}. These representations thus retain only one parameter per USPDF, and therefore become very noisy on small stellar samples.
	\item Histograms of mean (orange line), median or mode values of USPDFs that are smoothed with log(age) uncertainties as:
	\begin{equation}
	N_{unc}(\tau) = \sum_i w_i \mathcal{N}(\tau | \tau_i, \sigma_{\tau,i}),
	\end{equation}
	where $\sigma_{\tau,i}$ is the uncertainty in log(age) for USPDF $i$, $\mathcal{N}(\tau | m, v)$ is the PDF of the normal distribution with mean $m$ and standard deviation $v$ and the summation is done over all USPDFs for all stars in the survey. This representation is much smoother than histograms of mean, median or mode values, although it retains only two parameters ($\tau_i$ and $\sigma_{\tau,i}$) for each USPDF. Smoothed distribution of median and mode values are not shown in \autoref{fig:surveys} to avoid overloading the plot.
	\item Stacked fits to USPDFs (green dashed line in \autoref{fig:surveys}):
	\begin{equation}
	N_{fit}(\tau) = \sum_i w_i F_i(\tau, p_i), \label{eq:fit}
	\end{equation}
	where $F_i$ is the best-fitting function for a solution's PDF in log(age) and $p_i$ are the parameters of the fit. Again, the summation is done over all solutions (USPDFs) for all stars in the survey.
	\item ``Full PDF'' -- the sum of all PDFs for all stars in the survey (red line in \autoref{fig:surveys}) -- this sum is produced by UniDAM by directly adding up all PDFs for each solution in the fitting process. For each solution the detailed PDFs are however not stored, as it will require much more space than already used and will be difficult to work with. 
\end{itemize}

The last two representations save more detailed information about log(age) PDFs than histograms and smoothed histograms of single parameters, and hence are more helpful in recovering real underlying age distributions.

\autoref{fig:surveys} shows that stacked fits to USPDFs $N_{fit}(\tau)$ (see \autoref{eq:fit})
are close to the full PDF, which confirms that fits are a reliable proxy for the full PDF. This means, that for a subset of survey stars one can use stacked fits to USPDFs for stars in this subset in place of the full PDF for the same subset. The advantage is that UniDAM output allows to reconstruct the log(age) PDF for an arbitrary subset of stars without re-running the fit and without storing full PDF data for each star, which would take about 30 times more space. For that reason, in our subsequent analysis we use $N_{fit}(\tau)$.

From \autoref{fig:surveys}, it is also clear that different representation of the age distribution have a substantially different shape for most surveys. Thus, the choice of the representation might affect further conclusions based on the age distribution of stars. It is however unclear which distribution is closest to the real age distribution. As we show below with the use of simulated data for which the age distribution is known, none of them is in fact a good representation of the real age distribution, so none can be reliably used as an age tracer, at least not directly.

\begin{figure*}
    \myimageTwo{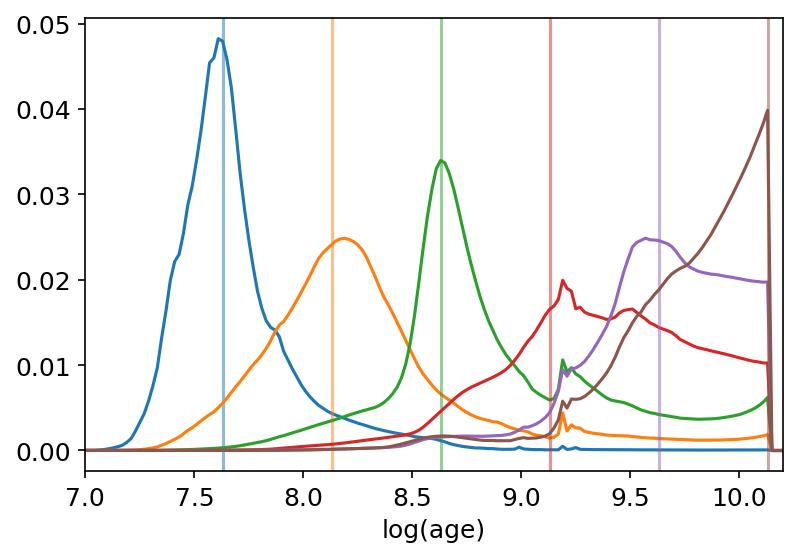}{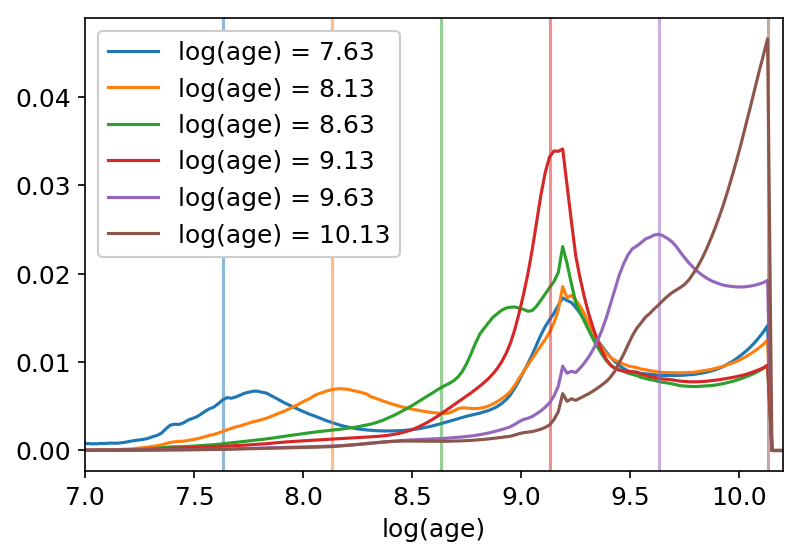}
    \caption{Examples of log(age) PDFs derived with UniDAM (same as red lines in \autoref{fig:surveys}) for simulated mono-age populations with $\log g$ and $\feh$\, distributions taken from APOGEE (left) and RAVE-on (right). Vertical lines indicate the true age for each population.}\label{fig:mock_pdfs}
\end{figure*}

To further illustrate the problem, we build an artificial survey using stellar isochrone models with a constant age $\tau_0$ and feed it into UniDAM. The real distribution in log(age) will be a delta function $F(\tau) = \delta(\tau - \tau_0)$. However, the log(age) PDF produced by UniDAM will not be a delta function. A few examples of these log(age) PDFs as produced by UniDAM are shown in \autoref{fig:mock_pdfs}. There we show log(age) PDFs for mono-age populations simulated in a way to mimic  APOGEE \citep{APOGEE} and RAVE-on \citep{2016arXiv160902914C} surveys (see \autoref{sec:build_mamc} for details on how the simulation were done). Strong difference in PDF shapes between APOGEE- and RAVE-on-based populations are due to different fractions of main-sequence stars in APOGEE and RAVE-on surveys: main sequence stars contribute more to the high-age part of the log(age) PDF. In most cases, log(age) PDFs have a peak near the population age, however the distributions are clearly non-Gaussian: they are very broad and in some cases show more than one peak.
This is not a mistake, but an inherent property of the Bayesian isochrone fitting method used in UniDAM. In fact, almost any other isochrone fitting method will have the same property: isochrones overlap on the \HRD, resulting in  the degenerate relation between observed spectroscopic parameters and physical properties of stars. 

We can try to reconstruct the real log(age) distribution from the observed log(age) PDF for a given stellar population. To do that, we need to simulate a set of mono-age mock catalogues (hereafter MAMC) and produce log(age) PDFs for each catalogue in this set. If we have an observed mono-age population (for example, an open cluster), we can compare its log(age) PDFs to the log(age) PDFs for MAMCs. The age for MAMC with the closest PDF will be the estimate of the age of the observed population. If the population consists of stars with a set of ages $\tau_1, \tau_2, ...\tau_n$, its log(age) PDF will be the sum of log(age) PDFs for MAMCs with same ages $\tau_1, \tau_2, ...\tau_n$. We can formulate an inverse problem and try for an arbitrary observed population to represent its log(age) PDF as a linear combination of the MAMC log(age) PDFs, with coefficients as a function of log(age) giving an estimate of the true age distribution in the observed population. The details of this approach are explained below.

\section{Inversion method}
\subsection{Main equation}
The idea of the  method proposed here is to try to represent the stacked log(age) PDF or analogous function for a given set of stars with the linear combination of log(age) PDFs for mono-age mock catalogues (MAMC). The method of MAMC construction is described below in \autoref{sec:build_mamc}. The important point here is that MAMC should be constructed in such a way that the distribution of its stars in $\log g$ and metallicity $\feh$ will be close to the distribution in these parameters for the considered set of stars.
In our approach MAMC with log(age) of $\tilde{\tau}$ contains stars with their spectrophotometric parameters taken from PARSEC model isochrones with the same log(age) $\tilde{\tau}$. Uncertainties were assigned to be close to observational ones (see more on how the uncertainties were chosen in \autoref{sec:build_mamc}). Because of the uncertainties in parameters, UniDAM will give for each star a log(age) PDF which will be different from a delta function $\delta(\tau - \tilde{\tau})$. In some cases, especially for stars on the main sequence, the PDF in log(age) will not have its peak around $\tilde{\tau}$, but will rather be close to an exponential function with its peak at the largest value of log(age) covered by models (see \autoref{fig:age_pdf_examples}). The exponential distribution in log(age) is equivalent to a flat distribution in linear age, which means that age is poorly constrained or unconstrained in these cases. A stacked PDF of a MAMC, similarly, does not necessarily have its maximum at $\tilde{\tau}$. In \autoref{fig:mock_pdfs} we show PDFs for several MAMC constructed to have their $\log g$ and $\feh$ distributions similar to those for APOGEE \citep{APOGEE} and RAVE-on \citep{2016arXiv160902914C} surveys.

We designate the log(age) PDF, as produced by UniDAM for a given survey, as $C(\tau)$. We aim at estimating the underlying log(age) distribution of stars $N(\tilde{\tau})$ for the same survey. For the MAMC with log(age) $\tilde{\tau}$ we designate the log(age) PDF as $P(\tau, \tilde{\tau})$. If the sizes of the survey and all MAMCs are infinite, the following equation should hold:
\begin{equation}
C(\tau) = \int_0^{\infty} P(\tau, \tilde{\tau}) N(\tilde{\tau}) d \tilde{\tau} \label{eq:diff}.
\end{equation}
Thus, to get the unknown age distribution $N$ from the observed PDF $C$ we need to solve the integral equation \ref{eq:diff}. We note however that in reality \autoref{eq:diff} holds only approximately, because the number of stars in the survey and in each MAMC is finite, thus both $C(\tau)$ and $P(\tau, \tilde{\tau})$ are subject to stochastic deviations. These deviations limit our ability to find the solution for \autoref{eq:diff}.

Let us consider as an example a case where the survey contains several mono-age populations (for example, a set of open clusters). This implies that $N(\tilde{\tau}) = \sum_i N_i \delta(\tilde{\tau} - \tau_i)$, with $N_i$ being the fractional weight of the $i$-th population, which has log(age) $\tau_i$. Following \autoref{eq:diff}, we get that $C(\tau) = \sum_i N_i P(\tau, \tau_i)$, in other words, log(age) PDF for such a survey is a weighted sum of PDFs of mono-age populations.
In fact, the PDF from UniDAM $C(\tau)$ is defined over the grid of log(ages) $(\tau_i, i = 1,..n)$. Similarly, MAMC PDFs $P(\tau, \tilde{\tau})$ are defined only for $\tau \in \tau_i$. Thus we can substitute the function $P(\tau, \tilde{\tau})$ with a quadratic matrix of the size $n\times n$ and functions $N(\tilde{\tau})$ and $C(\tau)$ with vectors $\vec{N}(\tilde{\tau})$ and $\vec{C}(\tau)$ of the size $n$, and \autoref{eq:diff} can be rewritten as a system of linear equations:
\begin{equation}
C_j = \sum_{i} P_{j, i} N_i.\label{eq:main}
\end{equation}
There is an obvious non-negativity constraint for this system: $N_i \ge 0$ for all values of $i$. Therefore, similarly to the example provided above, we try to represent the survey population as a superposition of $i$ mono-age populations with ages $\tau_i$. 

\subsection{Solving the main equation}\label{sec:method}
In the system of linear equations (\autoref{eq:main}), $C_j$ are taken from the survey log(age) PDF and $P_{j, i}$ comes from MAMC log(age) PDFs. We want to solve this system of linear equations for $N_i$, bearing in mind the non-negativity constraint. It is possible to find a solution by means of non-negative least squares method \citep[NNLS, see][]{NNLS}, which maximizes the following function:
\begin{equation}
L_0 = - \sum_{j=1}^{n} (C_j - \sum_{i=1}^{n} P_{ji} N_i)^2 .
\end{equation}
This will typically lead to a result in which only several components of $\vec{N}$ will be non-zero, which is not physical -- we expect a rather smooth log(age) distribution.

In order to get a smooth result we use regularized likelihood maximization. We chose a Tikhonov regularization, that favours smoother solutions by adding a sum of squares of the solution's first derivatives\footnote{We performed tests with a version of Tikhonov regularization where the sum of squares of the solution values, rather than derivatives, is used, which resulted in very similar however slightly noisier results.}.
We thus maximize the following function:
\begin{equation}
L = - \sum_{j=1}^{n} (C_j - \sum_{i=1}^{n} P_{ji} N_i)^2 - \lambda \sum_{i=1}^{n} \left( \frac{d N_i}{d \tau} \right)^2, \label{eq:likelihood}
\end{equation}
where $\lambda$ is a regularization parameter. Solutions with narrow spikes will produce large absolute values of $ \frac{d N_i}{d \tau}$ and will have smaller values of likelihood function $L$, as compared to smoother solutions, even if the latter produce larger differences between $C_j$ and $ \sum_{i=1}^{n} P_{ji} N_i$.

We can use finite difference formulas to calculate $\frac{d N_i}{d \tau}$ from $N_i$, and rewrite:
\begin{equation}
	\frac{d N_i}{d \tau } = \mathcal{T} \vec{N},
\end{equation}
where, $\mathcal{T}$ is a Toeplitz matrix representation of the first derivative:
\begin{equation}
\tiny{
	\mathcal{T} = \begin{pmatrix}
	-1.5 & 2.   & -0.5 & 0    & \cdots & \cdots & \cdots & 0\\
	0    & -1.5 & 2.   & -0.5 & 0      & \cdots &  \cdots & 0\\
	0    & 0    & -1.5 & 2.   & -0.5   & 0      &  \cdots & 0\\
	\vdots & \ddots & \ddots & \ddots & \ddots & \ddots & & \vdots\\
	\vdots &        & \ddots & \ddots & \ddots & \ddots & \ddots &\vdots\\
	0 & \cdots &  0      & 0    & -1.5 & 2.   & -0.5 & 0      \\
	0 & \cdots &  \cdots & 0    &  0 & -1.5 & 2.   & -0.5      \\
	0 & \cdots &  \cdots & 0 & 0.5 & -2 & 1.5 & 0 \\
	0 & \cdots &  \cdots & \cdots & 0 & 0.5 & -2 & 1.5 \\
	\end{pmatrix}
}
\end{equation}
In this matrix, the $i$-th row represents coefficients for the 2nd order forward formula for the first derivative:
\begin{equation}
f'(x_i) = - \frac{3}{2} f(x_i)  + 2 f(x_{i+1}) - \frac{1}{2} f(x_{i + 2}),
\end{equation}
with an exception of the last two rows, that represent similar coefficients for the backward formula:
\begin{equation}
f'(x_i) = \frac{1}{2} f(x_{i - 2}) - 2 f(x_{i-1}) + \frac{3}{2} f(x_i)
\end{equation}

Thus, \autoref{eq:main} can be rewritten as:
\begin{equation}
\vec{C}' = P' \vec{N}\label{eq:main2},
\end{equation}
where:
\begin{eqnarray}
P' &=& \begin{pmatrix}
P \\
\lambda \mathcal{T} 
\end{pmatrix} \\    
\vec{C}' &=& (\vec{C} | 0 \cdots 0)^T
\end{eqnarray}

In this notation, maximization of $L$ from \autoref{eq:likelihood} is equivalent to minimization of $\begin{Vmatrix}\vec{C}' - P'\vec{N}\end{Vmatrix}_2$, with a constraint that all components of $\vec{N}$ are non-negative. Again, we can use NNLS to obtain the optimal vector $\vec{N}$. The solution can be characterised by the sum of residuals $R = \begin{Vmatrix}\vec{C} - P \vec{N}\end{Vmatrix}_2$, which indicates, how close the log(age) PDF for the derived age distribution is to $C(\tau)$.

\subsection{Searching for optimal $\lambda$ value} \label{sec:lambda}
The problem of the regularization approach is that the parameter $\lambda$ has to be properly chosen in order to obtain the correct solution of the problem. Two extreme cases are $\lambda = 0$ and $\lambda = \infty$. In the first case no regularization is active, and \autoref{eq:main2} is reduced to \autoref{eq:main}. In that case the solution is most precise however not smooth. If $\lambda = \infty$ the regularization dominates the solution, such that $N(\tau)$ becomes a constant, equal to the mean value of $C(\tau)$ -- this is the smoothest possible solution, which is very imprecise. In this section we describe an empirical method of obtaining the value of $\lambda$ that gives the result we consider to be optimal.

First of all, we note that both $P$ and $\vec{C}$ are subject to statistical variations due to the limited number of sources in both the observed survey and in the simulated MAMC. Because each star can contribute to a wide range of log(age) values, variations of the PDF at different values of log(age) $\tau$ are not independent. In order to account for this effect properly, we chose to produce five realizations of MAMC (differing only by random number algorithm seed) and to split the observed survey into five parts. This gives us five $P$ matrices $P_p, p = 1, 2,.. 5$ and five $\vec{C}$ vectors $\vec{C}_q, q = 1, 2,.. 5$. Using this information, we can make use of the cross-validation technique, requiring that the solution obtained with one combination of $p$ and $q$ values should be good for all other combinations. 

For a given value of regularization parameter $\lambda$ we can thus build a set of 25 equations similar to \autoref{eq:main2}, with all possible combinations of $\vec{P}_p$ and $\vec{C}_q$. Let us now focus on the combination with $p = p_0$ and $q = q_0$. For this combination, we write \autoref{eq:main2} as:
\begin{equation}
    \vec{C}'_{q_0} = \vec{P}'_{p_0} \vec{N}, \label{eq:main3}
\end{equation}
 and designate a solution of this equation as $\vec{N}_{p_0, q_0}$. We can further build a matrix of residuals with respect to all combinations of $\vec{C}$ and $\vec{P}$:
\begin{equation}
   R_{m,n, p_0, q_0} = \begin{Vmatrix}\vec{C}_n - \vec{P}_m \vec{N}_{p_0, q_0}\end{Vmatrix}_2. \label{eq:r_mn}
\end{equation} 
As our quality parameter we take the following expression:
\begin{equation}
  Q_{p_0, q_0} = \frac{1}{24} \left(\sum_p \sum_q R_{m, n, p_0, q_0} - R_{p_0, q_0, p_0, q_0}\right). \label{eq:q}
\end{equation}

\begin{figure}
    \myimage{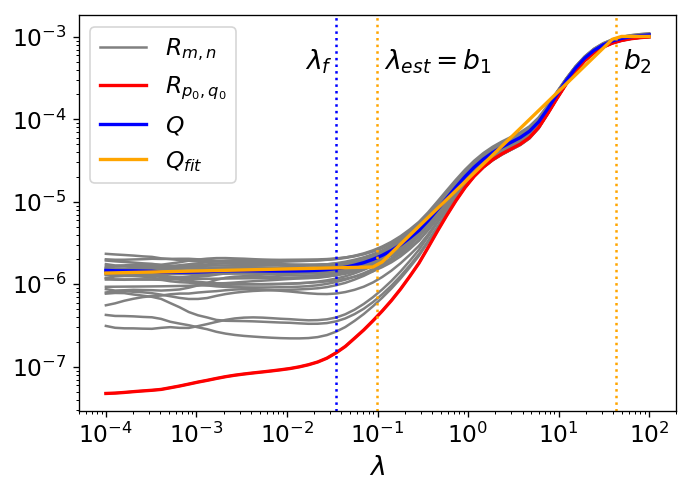}
    \caption{An illustration for the choice of optimal $\lambda$. The plot shows all residual functions $R_{m, n}$ (grey lines, see \autoref{eq:r_mn}) and highlights the $R_{p_0, q_0}$ in red and $Q$ (\autoref{eq:q}) in blue. The fitted piecewise linear function $Q_{fit}$ is shown in orange (see \autoref{eq:qfit}). Vertical lines indicate the turning points of the fitted piecewise linear function $Q_{fit}$ (orange dashed lines) and the finally adopted $\lambda$ value at the intersection point of $Q$ and $Q_{fit}$ (blue dashed line).
    }\label{fig:optimal}
\end{figure}

 All $R_{m,n,p_0, q_0}$ and ${Q_{p_0, q_0}}$ are functions of the regularization parameter $\lambda$.
 We show an example of $R_{m,n,p_0, q_0}(\lambda)$ and $Q_{p_0, q_0}(\lambda)$ functions in \autoref{fig:optimal}. $R_{p_0, q_0, p_0, q_0}$ decreases as $\lambda$ decreases -- with less regularization it is possible to have more accurate solution of \autoref{eq:main2}. For values of $R_{m,n,p_0, q_0}$ for $m \neq p_0$ or $n \neq q_0$ the decreasing trend with decreasing $\lambda$ stops or even reverses at some point. This happens because solutions $\vec{N}_{p_0, q_0}$ of \autoref{eq:main2} that are accurate for the combination $\vec{P}_{p_0}$ and $\vec{C}_{q_0}$ become too ``specialized'', and are not as accurate for other combinations of $\vec{P}_p$ and $\vec{C}_q$. As an optimal value of $\lambda$ we will take the point where the rapid decrease of $Q_{p_0, q_0}$ (hereafter labelled simply as $Q$) with decreasing $\lambda$ stops or slows down. To find this point, we first fit a following piecewise linear function to the $Q$-function in log-log space:
\begin{equation}
    Q_{fit}(\log\lambda) = \left\{ \begin{array}{l}
        a + c (\log\lambda - \log b_1), \textrm{if}\, \lambda \le b_1 \\
        a + d (\log\lambda - \log b_1), \textrm{if}\, b_1 < \lambda \le b_2 \\
        a + d (\log b_2 - \log b_1),\textrm{if}\, \lambda > b_2
    \end{array}\right. \label{eq:qfit}.
\end{equation}

 Function $Q_{fit}$ has three linear segments, with transitions at values of $\lambda = b_1$ and $\lambda = b_2$. Slopes are $c$ in the first segment (close to zero in \autoref{fig:optimal}), $d$ in the second segment and zero in the third segment. Parameter $a$ represents the value of $Q_{fit}(b_1)$. This choice of the representation for the piecewise linear function is motivated by the ease of the initial guess for the fitted parameters. All five parameters ($a, b_1, b_2, c, d$) are fitted simultaneously.

As the first estimate for the optimal regularization parameter value we take the first turning point of the fitted function $\lambda_{est} = b_1$. In many cases the turn of the $Q$ function is not as sharp as the one of the fitted function, and $\lambda_{est}$ is an overestimation of the optimal value of $\lambda$. To correct for that, we take as our final estimate $\lambda_{f}$ the first point to the left of $\lambda_{est}$ where $Q$ and $Q_{fit}$ intersect, as shown in \autoref{fig:optimal}. 
This estimate in some cases tends to produce a noisier ``over-fitted'' result, while the $\lambda_{est}$ tends to produce smoother ``under-fitted'' one. On the average, however, $\lambda_{f}$ produces better results than $\lambda_{est}$.
 Our choice of the $\lambda_{f}$ is further validated with tests described below in \autoref{sec:tests}.
 
We can repeat the above procedure for every possible combination of $p_0$ and $q_0$, thus obtaining 25 different values of $\lambda_{f}$ and corresponding solutions $\vec{N}_{p_0, q_0}$ of \autoref{eq:main3}. As the final solution we can take the mean of $\vec{N}_{p_0, q_0}$, and as a measure of the uncertainty -- their standard deviation. For a more detailed uncertainty analysis see \autoref{sec:uncertainty}.

\section{Mono-age mocks construction}\label{sec:build_mamc}
In order to properly recover the age distribution for a given survey, which we call a base survey, we need to construct MAMCs in such a way, that the distribution of stars in each MAMC in physical parameters and their uncertainties will be close to that of the base survey, at the same time retaining age information. To achieve that, we follow the procedure below. 

We build MAMC using PARSEC models \citep{PARSEC} to simulate stars. 
In order to mimic the random observational scatter we take instead of each model a sample of 25 models with random normally distributed perturbations added in $\teff, \log g$ and $\feh$.
To assign a proper amplitude of perturbations, we randomly select uncertainties $\sigma_P =  \{ \sigma_T, \sigma_{\log g}, \sigma_{\feh} \}$ from the base survey data for stars that have $P = \{\teff, \log g\,\textrm{and}\,\feh \}$ close to the model, such that the difference between physical parameters for the model and for the base survey star is smaller than the mean uncertainty in the respective parameter for the base survey $P_{model} - P_{star} < \textrm{mean}(\sigma_P)$. This provides a way to reconstruct the scatter and the systematic variations of $\sigma_P$ across the parameter space. Selected uncertainties are not only used as perturbation amplitudes, they are also assigned as ``observational'' uncertainties for models.
The photometric magnitudes of all models are perturbed with a Gaussian noise with a scale of $0.^m025$, to reproduce typical photometric uncertainties in 2MASS and AllWISE.

The set of models is then truncated in the $\teff, \log g, \feh$ space to the footprint of the base survey in that space, models outside of that footprint are excluded. In order to build MAMC from that set of models, we sample a pre-defined number of models $\nmock$ as follows.
We bin stars from the base survey in $\log g - \feh$ space.
The fraction of models sampled for a given MAMC from each bin is equal to the fraction of base survey stars in that bin. Within each bin, models are selected randomly with probabilities proportional to the fraction of the initial mass function represented by that model. 

We cannot use all three physical parameters ($\teff, \log g, \feh$) for binning, as in this case the resulting MAMC will be too close to the base survey itself, retaining almost no information about the MAMC age value. On the other hand, using just one parameter will produce MAMC that will not resemble the base survey, making inversion impossible. We decided to use as a first parameter for the binning $\feh$, as it is the only parameter out of the three that is independent of stellar mass, and as such is expected to be the same for all stars of the same origin. The second parameter is $\log g$, as it helps to distinguish main sequence stars and giants, that have very different contribution the log(age) PDF. Such distinction is impossible to make with $\teff$.

The above procedure is repeated for each value of log(age) $\tau$ over the considered grid: $6.61 \leq \tau \leq 10.13$\,dex with a step of $0.02$\,dex or age between approximately $4\cdot 10^6$ and $13.5\cdot10^9$ years. As a result we get a set of MAMCs that is fed into UniDAM to obtain a set of log(age) PDFs $P_{j, i}$.

\section{Tests with mock data}\label{sec:tests}
In this section, we use simulated data to validate the choice of the optimum regularization parameter $\lambda$ and to determine the influence of the survey size $\nsim$ and MAMC size $\nmock$ on the precision and the accuracy of the inversion result. We simulate a survey with pre-defined age distribution $\vec{N}_{input}$ as a concatenation of MAMC simulated in a way described in \autoref{sec:build_mamc}, with the size of $i$-th MAMC defined by $N_{input, i}$. We than obtain with UniDAM the log(age) PDF for the simulated survey and apply the inversion as described in the \autoref{sec:method}. We can than compare the result of the inversion with the input age distribution, to get an estimate of the accuracy and precision of the method.
Below we describe the process in detail.

\subsection{Input distributions}
\begin{figure*}
    \myimage{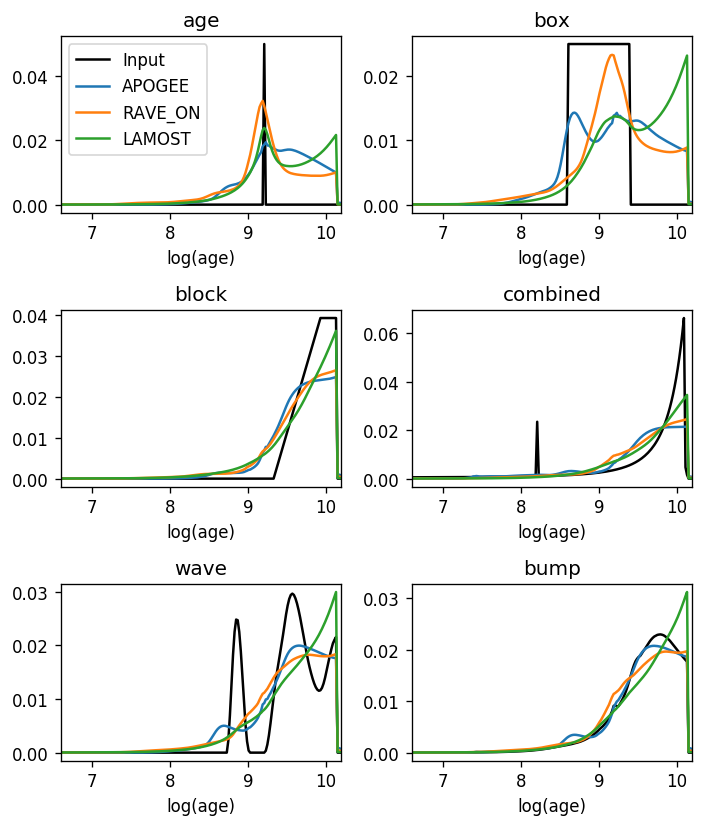}
    \caption{Input log(age) distributions and log(age) PDFs produced with UniDAM for different base surveys. Plots are for six cases considered in this work. For the ``age'' input the true PDF was re-scaled to make other lines visible.}\label{fig:inputs}
\end{figure*}

For this work we chose six input age distributions, aiming to emulate critical as well as more common cases. Every distribution was generated using one of the three base surveys: APOGEE DR14 \citep{APOGEE}, LAMOST DR3 \citep{LAMOST} and RAVE-on \citep{2016arXiv160902914C}. In \autoref{fig:inputs} we show the input log(age) distributions and PDFs produced by UniDAM for the three base surveys. ``Age'' and ``box'' inputs implement a spike and step function. Such functions are hard to reproduce with our method, as the uncertainties in age determination tend to smooth out the distribution. ``Block'' and ``bump'' inputs show the opposite case of smooth slowly changing PDFs. Hence, we expect that the inversion will work best for them. ``Combined'' and ``wave'' inputs represent intermediate cases. Interestingly, log(age) PDFs for ``bump'' and ``wave'' simulations have very similar
shapes for all three base surveys, despite the very different input log(age) distributions. Note that the choice of base survey affects the log(age) PDFs generated by UniDAM. Most importantly, the log(age) PDF is typically not close to the true age distribution, with the only exception of the APOGEE result for the ``bump'' distribution.

\subsection{Test results}
\begin{figure*}
	\myimageFour{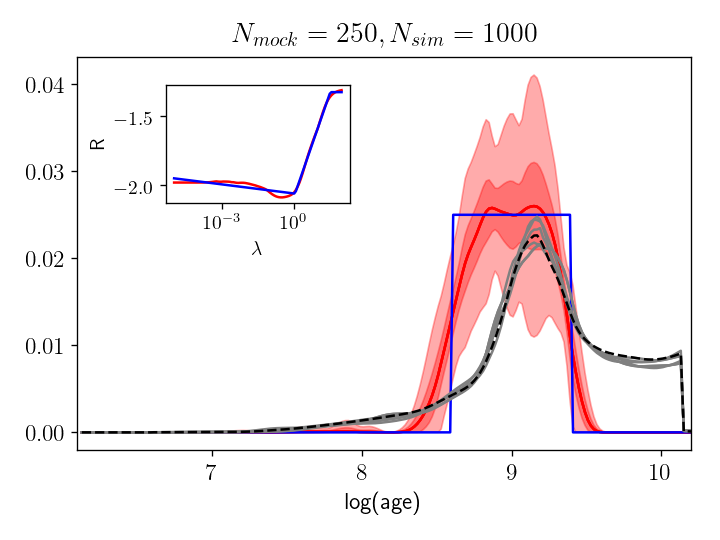}{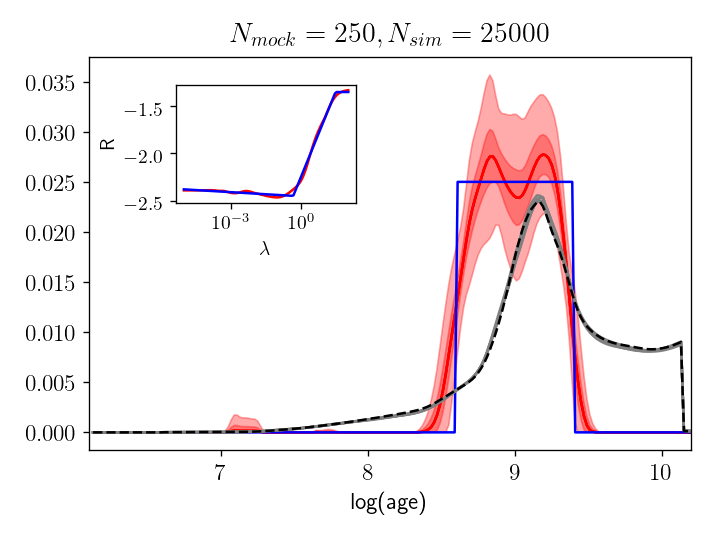}{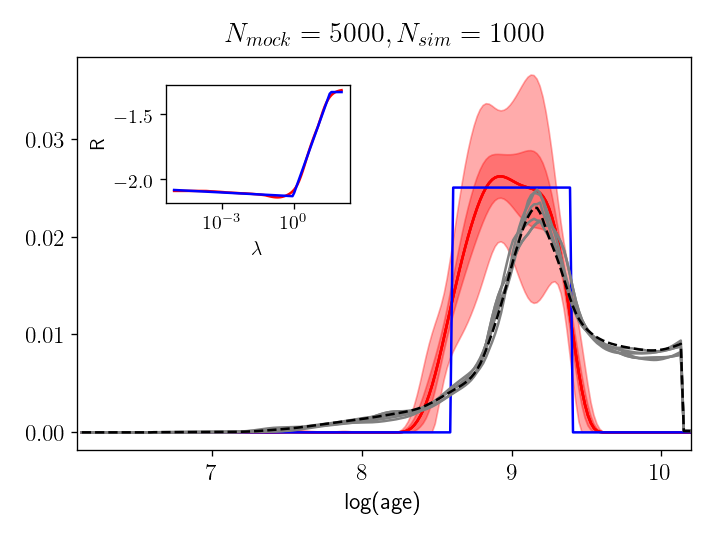}{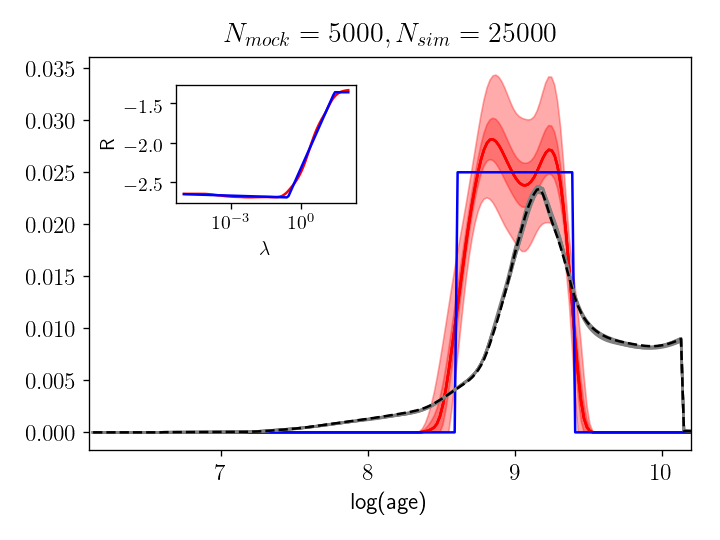}
	\caption{Results of the age inversion for ``box'' test data. Upper panels are for smaller MAMC ($\nmock = 250$), lower panels are for larger MAMC ($\nmock = 5000$). Left panels are for smaller simulated survey ($\nsim = 1000$), right panels are for larger simulated survey ($\nsim = 25000$). The solid blue line shows the input log(age) distribution, the solid red line is the result of the inversion with 68- (dark shading) and 95 (light shading) percent confidence intervals. Input log(age) PDFs for five realizations of simulated catalogue are plotted with grey lines and the black dashed line shows the log(age) PDF inferred from the inversion result. The inset in each panel shows the cross-validation curves (red) and piecewise-linear fit (blue) for one realization of the simulated catalogue (see \autoref{sec:lambda} for the description of the cross-validation curve).}\label{fig:test_example}
\end{figure*}

\begin{figure*}
	\myimageFour{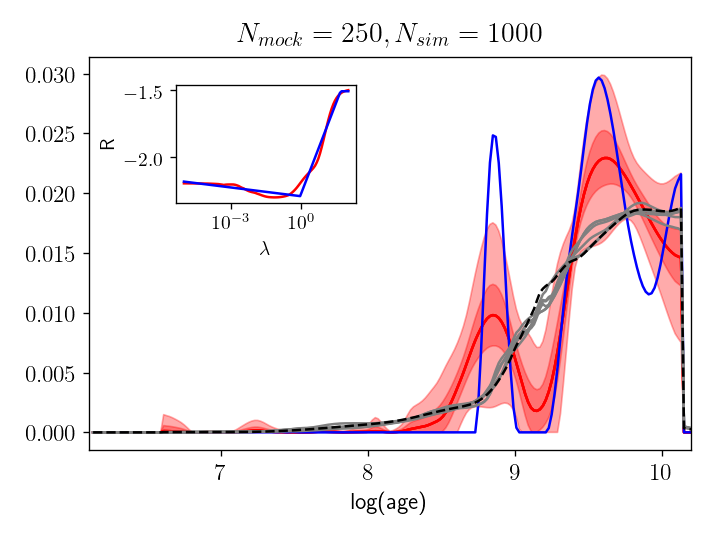}{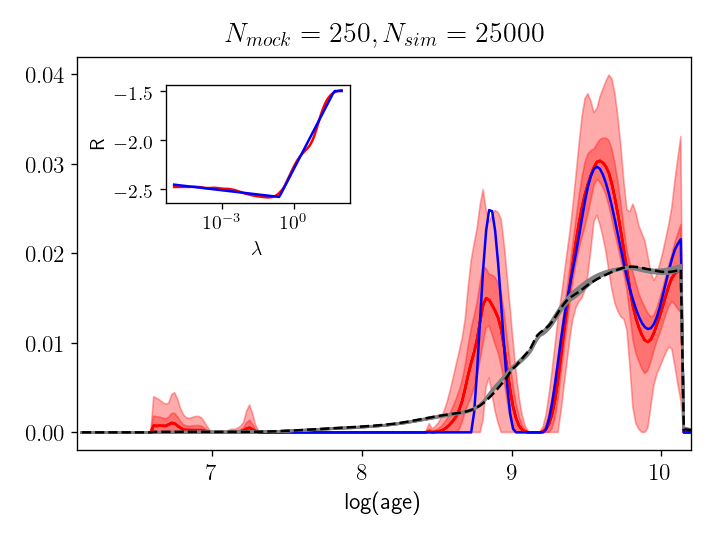}{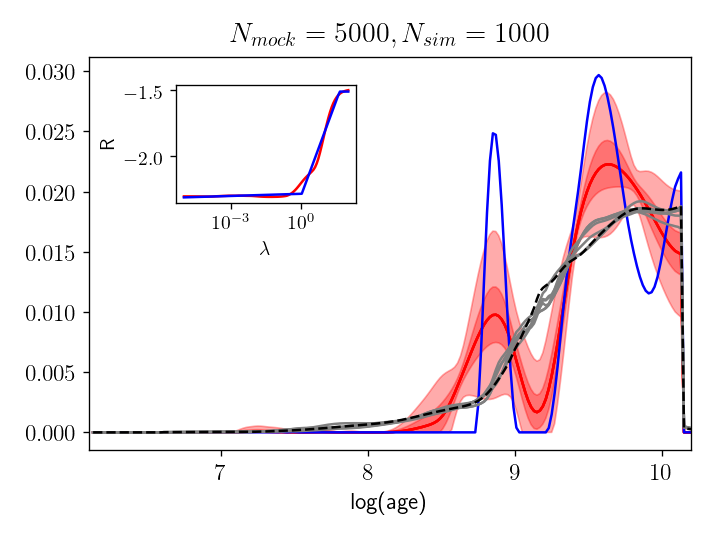}{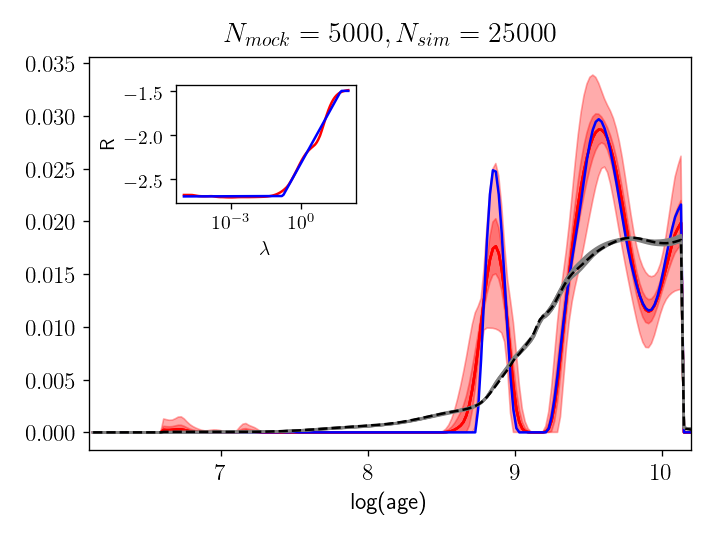}
	\caption{Same as \autoref{fig:test_example}, now for ``wave'' input.}\label{fig:test_example2}
\end{figure*}

In Figs. \ref{fig:test_example} and \ref{fig:test_example2} we show examples of test results, with different input distributions and varying $n_{sim}$ and $n_{mock}$. There we compare input distributions (blue lines) with age inversion results (red lines), and ``observed'' log(age) PDFs (dashed black lines) with log(age) PDFs as predicted by age inversion (grey lines, one for each combination of simulated survey and MAMC realizations).

Figs. \ref{fig:test_example} and \ref{fig:test_example2} illustrate that the result of the inversion is almost insensitive to $\nmock$ for small values of $\nsim$, as expected (see discussion in \autoref{sec:uncertainty}). On the other hand, for $\nsim \geq 25000$, MAMC size $\nmock$ plays a larger role. It is also clear that an increase in $\nsim$ and $\nmock$ increases the sensitivity of the inversion to rapid changes in log(age) distribution. For example, in the case of ``wave'' test, $\nmock = 250$ and $\nsim = 25000$ is enough to properly recover the second peak of log(age) distribution at $\tau \approx 9.6$\,dex, though we need to increase $\nmock$ to 5000 in order to properly recover the first, much narrower peak at $\tau \approx 8.9$\,dex.
For all simulations, ``observed'' (black lines) and predicted (grey lines) log(age) PDFs become almost indistinguishable for $\nmock = 5000$ and $\nsim = 25000$. 

\subsection{Uncertainty analysis}\label{sec:uncertainty}
An important part of every scientific result is a proper uncertainty. We can use our tests to provide a way to estimate uncertainties for age inversion results.

In \autoref{sec:lambda} we gave a method of obtaining the optimal value of regularization parameter $\lambda$. This parameter and a corresponding solution of \autoref{eq:main2} can be obtained for every combination of the five survey log(age) PDFs $\vec{C}_q$ and five MAMC log(age) PDF sets $\vec{P}_p$. Hence we can have 25 different solutions of \autoref{eq:main3}, with there differences being due to statistical variations in $\vec{P}$ and $\vec{C}$. 
Variations between these solutions can be used to estimate the uncertainty of the solutions. Because of the smoothness of the solution, which is imposed by the regularization, and because of the similarity between log(age) PDFs for MAMCs with similar log(age) values, differences between solutions at different values of log(age) $\tau$ are highly correlated. For these reasons, the distribution of differences between the inversion result and the true underlying age distribution will not be a Gaussian but a heavy-tailed distribution.

\begin{figure}
	\myimage{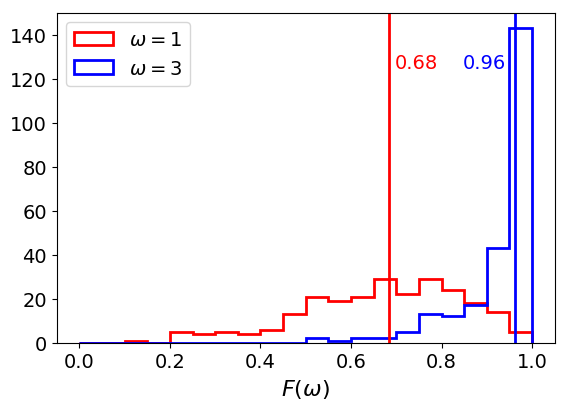}
	\caption{Distributions of the fraction $F(\omega)$ of values of $\tau$ with true error $\delta(\tau) < \omega \sigma(\tau)$, where $\sigma(\tau)$ is the estimated uncertainty. Shown are distributions for $\omega=1$ and $\omega=3$ for simulated data. Vertical lines and numbers represent median values of each distribution.}\label{fig:uncertainty}
\end{figure}

In the case of the tests that we are performing, we also use the knowledge of the true age distribution to find the relation between the variation between solutions $\sigma(\tau)$ and the true error $\delta(\tau)$  (the true error is defined as the absolute difference between input and output log(age) distributions). 
The fraction $F(\omega)$ of values of $\tau$ with $\delta(\tau) < \omega \sigma(\tau)$ can be calculated for each result. In \autoref{fig:uncertainty} we show distributions of $F(\omega)$ for $\omega=1$ and $\omega=3$. If $\sigma(\tau)$ is a correctly determined Gaussian uncertainty, than $F(1) = 0.68$ and $F(3) = 0.997$, as expected for 1- and 3-sigma confidence intervals. Measured $F(1)$ values for all our tests show a broad distribution with a mean of 0.67 and median of 0.685 -- very close to expectations. Measured $F(3)$ values also have a broad distribution, with a mean of 0.93 and median of 0.96 -- lower than the expected value, making $3 \sigma(\tau)$ effectively a 2-sigma rather than 3-sigma confidence interval estimate. This is because uncertainties have a heavy-tail non-Gaussian distribution. There are no clear trends visible for $F(\omega)$ values with $\nsim$ and $\nmock$, at least within the considered ranges of these parameters.

We note here, that the scatter of $F(\omega)$ around the median values is high, which means that values of $\sigma(\tau)$ can be seen only as an approximation of the uncertainty. 

One would expect the fractional uncertainty values to decrease as $(\nsim \nmock)^{-1/2}$. In reality, there relation is not as strong, because of complex correlations between the real age distribution and the corresponding log(age) PDF. Our estimate shows that the fractional uncertainty scales as $\nsim^{-0.4} \nmock^{-0.15}$. This means that one has to increase $\nmock$ by an order of magnitude to improve the result by about $30 \%$. Importantly, for a given $\nmock$ there seem to exist a maximum value of $\nsim$, above which the fractional uncertainty does not decrease with $\nsim$ at all. This is caused by the fact that in that regime fractional uncertainty becomes dominated by variations between realizations of MAMC, and not by the survey. The opposite is also true -- for a given $\nsim$ there is a maximum value of $\nmock$, above which the fractional uncertainty does not decrease, as it is dominated by survey PDF variations. The exact value of the maximum of $\nmock$ for a given $\nsim$ depends on the base survey and on the shape of the underlying age distribution and can vary by over an order of magnitude. The maximum considered value $\nmock = 5\,000$ is sufficient for surveys as large as $\nsim = 50\,000$. This is still below typical sizes of spectroscopic surveys. However, if we want to slice the survey into parts to trace, for example, age-metallicity relations, as is done below in \autoref{sec:monometallicity}, the size of each part will be on the order of 
$\nsim = 50\,000$ or even smaller. Even more importantly, all these measurements are done for artificial data, where systematic uncertainties are zero by definition, as simulated and mock catalogues are created from the same set of models. For real data, as we will show below, systematics will likely be the dominating source of uncertainty.

\section{Real data applications}

\subsection{Inversion of full surveys}\label{sec:full_survey}
\begin{figure*}
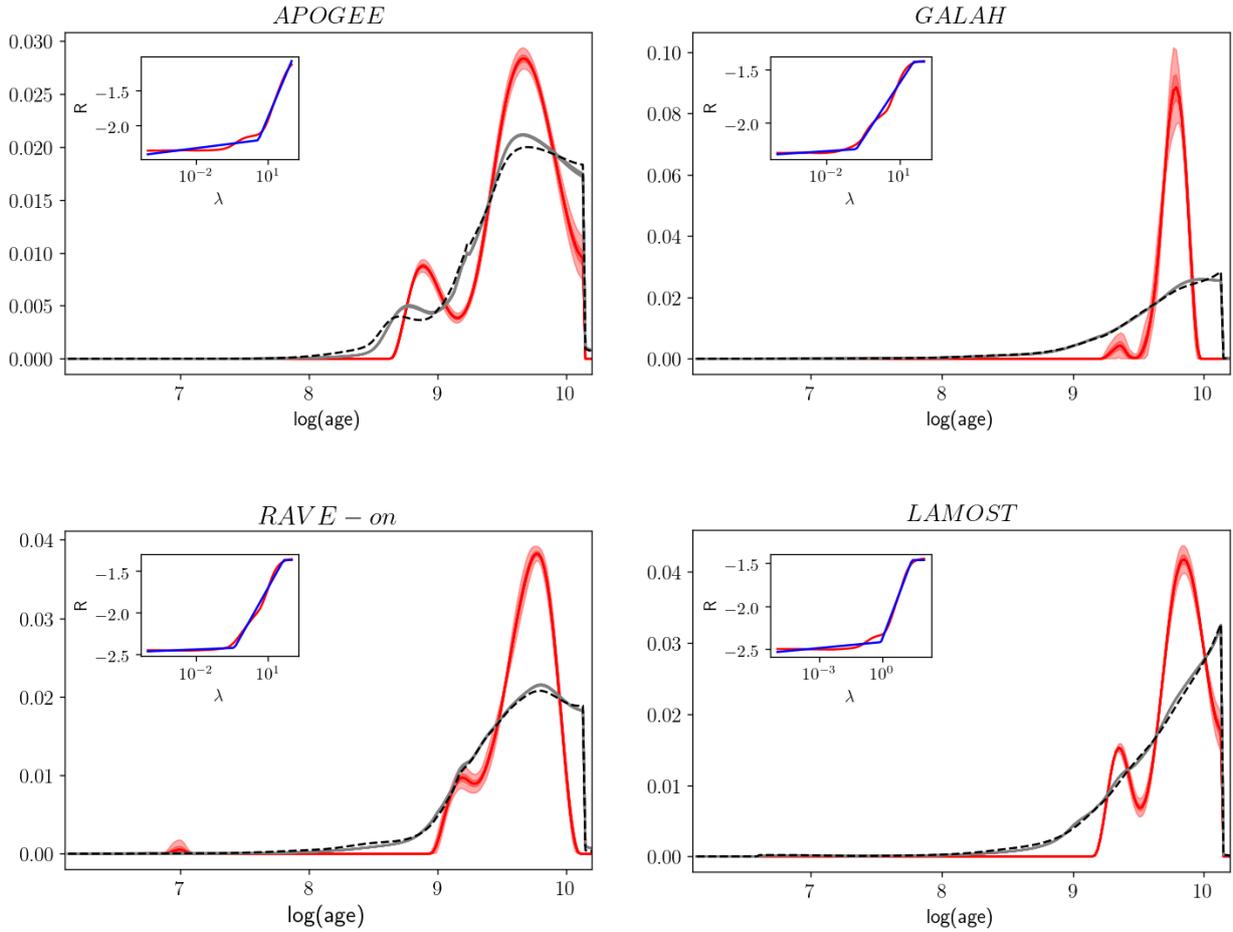

	\myimageFour{{run_APOGEE.json}.png}{{run_GALAH.json}.png}{{run_RAVE_ON.json}.png}{{run_LAMOST.json}.png}
	\caption{Results of age inversion for several surveys. The solid red line is the computed underlying age distribution with 68- (dark shading) and 95 (light shading) percent confidence intervals. Input log(age) PDFs for five parts of each survey are plotted with grey lines and the black dashed line shows the log(age) PDF inferred from the inversion result. The inset in each panel shows the cross-validation curves (red) and piecewise-linear fit (blue) for one part of the survey (see \autoref{sec:lambda} for the description of the cross-validation curve).}\label{fig:results_full}
\end{figure*}

We apply the inversion method described above to several large spectroscopic surveys, namely APOGEE \citep{APOGEE}, GALAH \citep{GALAH}, RAVE-on \citep{2016arXiv160902914C} and LAMOST \citep{LAMOST}. The results are presented in  \autoref{fig:results_full}. These plots indicate that all surveys have a bimodal age distribution, although the location, width and relative amplitudes for the two modes are different. 
There are several possible explanations for this fact.
The first possibility 
is that the trend is real and is caused by the target selection in the survey, with stars of different metallicity being observed in a different parts of the Galaxy. 
The second possibility is that one of the implicit assumptions of the method does not hold, namely, that in a given sample age does not depend on $\log g$ or $\feh$. Within the metallicity bin for a magnitude limited survey, observed stars with lower values of $\log g$, and thus with higher luminosities, are on the average located at larger distances than those with higher $\log g$, and thus with lower luminosities. Therefore age distribution might vary over the observed $\log g$ range. This may cause the inversion method to give incorrect results. It remains unclear however, which of the effects dominates. 

Last but not least, PARSEC models and spectroscopic measurements can have a systematic offset between them, which can cause a complex systematic age bias, which cannot be accounted for in modelling. 

  
  Possible systematic effects manifest themself also in the difference between the inversion result and log(age) PDFs of the survey, indicated in \autoref{fig:results_full} with black dashed and grey solid lines. This difference is considerably larger than the one we obtain for the simulated data, where the systematic offset is zero, even though the survey size $\nsim$ is larger than those considered in the simulation. 

\subsection{Inversion of mono-metallicity populations}\label{sec:monometallicity}
The limitation of the inversion of the full survey is that the age distribution is assumed to be the same for all distances and all metallicities covered by the survey, which in general does not hold, as more metal poor stars tend to be older.
In order to mitigate this problem we made a separate study of RAVE-on, LAMOST \citep{LAMOST} DR4 and APOGEE stars, splitting them in several metallicity bins and applying age inversion to each bin separately. 

\begin{figure}
	\myimage{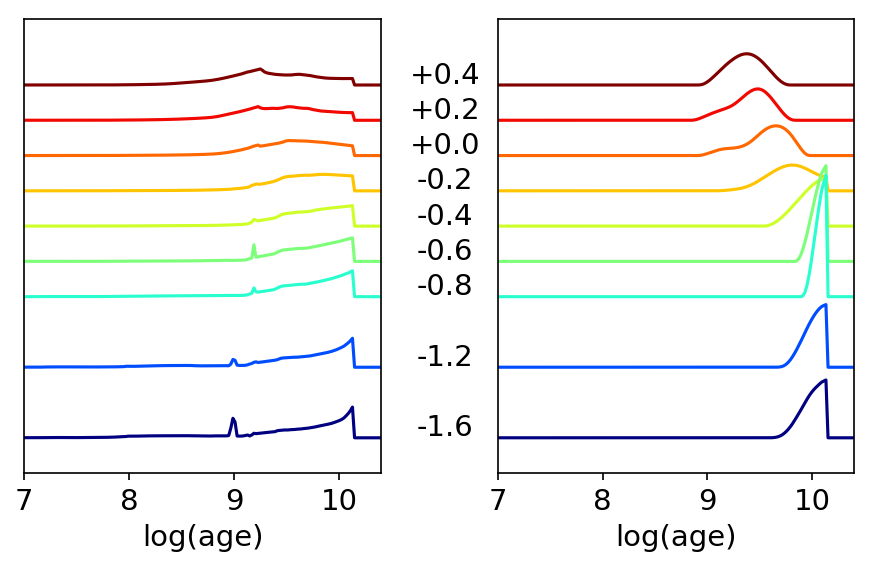}
	\caption{Results of age inversion for RAVE-on metallicity slices: log(age) PDFs as produced by UniDAM (left) and underlying age distribution computed through the inversion (right). Plots for various metallicities are offset in vertical direction for visual purposes, with numbers  between plots indicating \feh\, values.}\label{fig:results_rave_on}
\end{figure}

In \autoref{fig:results_rave_on} we present the inputs (log(age) PDFs) and outputs (underlying age distributions) of the age inversion procedure for RAVE-on stars. Inputs (log(age) PDFs) are typically broad and for metal-poor populations show an extra narrow peak at log(age) $\tau \approx 9$, which is associated with red-clump stars. The underlying age distributions computed through the inversion are a lot smoother and narrower, with no secondary peaks. The peak of the underlying age distribution goes from log(age) $\tau = 9.37$ for highest metallicity bin $\feh = 0.4$\,dex to the maximum possible log(age) of $\tau = 10.13$ for  $\feh < -0.4$\,dex. At very low metallicities $\feh \le -1.2$ a gradual broadening of the underlying age distribution is observed, which is caused by the presence of a lower number of stars in metal-poor bins and thus a reduced age resolution (see test results with low $\nsim$ value in \autoref{fig:test_example} and \ref{fig:test_example2}).

\begin{figure}
	\myimage{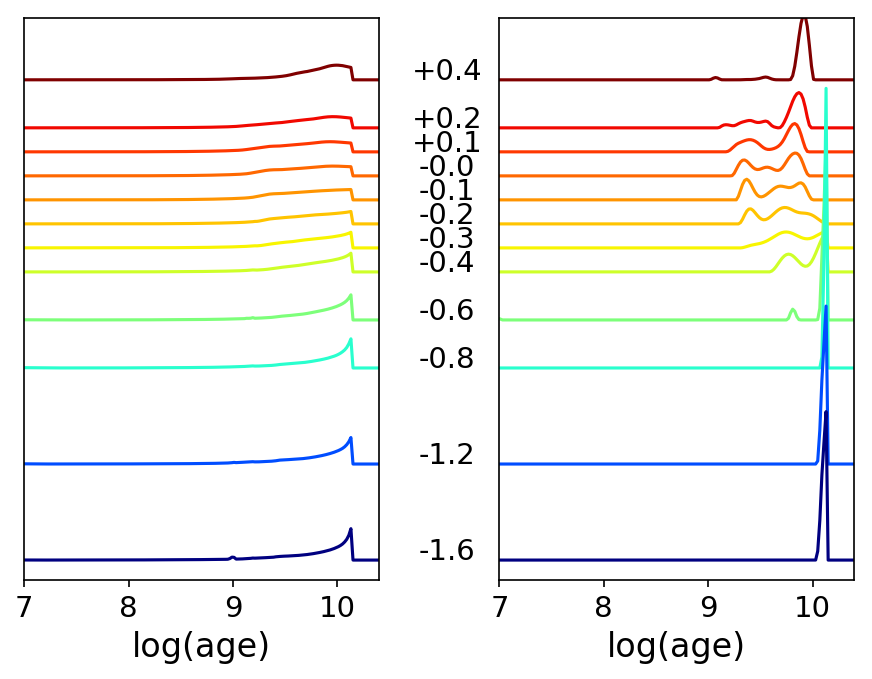}
	\caption{Same as \autoref{fig:results_rave_on}, now for the LAMOST survey.}\label{fig:results_lamost}
\end{figure}

\begin{figure}
	\myimage{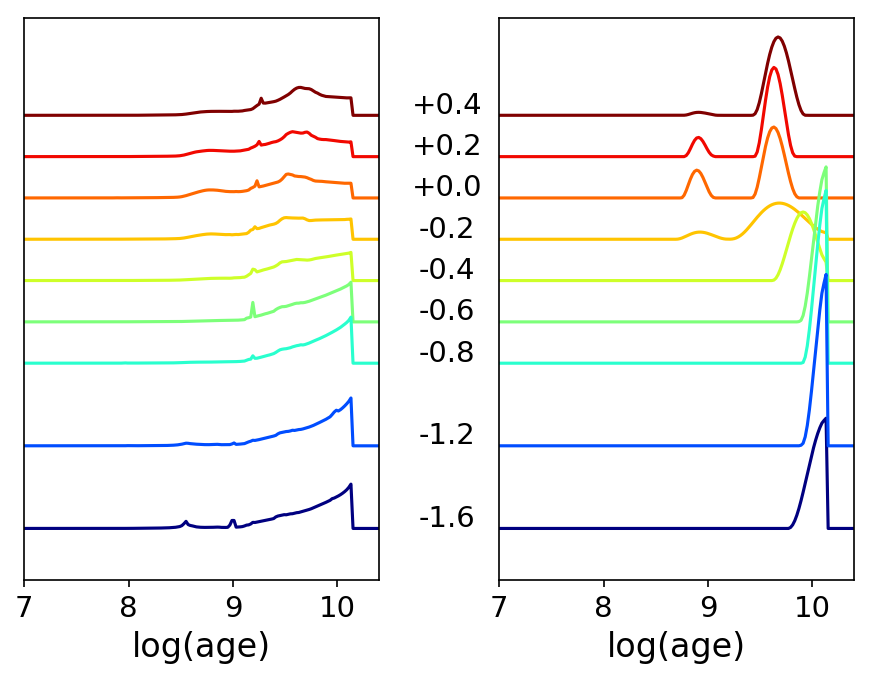}
	\caption{Same as \autoref{fig:results_rave_on}, now for the APOGEE survey.}\label{fig:results_apogee}
\end{figure}

For LAMOST, we use a constraint of $\teff < 7000$K in this work, as there are visible pipeline artefacts in the \HRD\ at higher temperatures, which we are not able to simulate. Even with that cut, LAMOST contains about an order of magnitude more stars than RAVE-on or APOGEE. 
The results of the age inversions for the LAMOST sample are shown in \autoref{fig:results_lamost} and for the APOGEE sample in \autoref{fig:results_apogee}. Results for LAMOST show narrower underlying age distributions than RAVE-on or APOGEE results, likely due to larger statistics and hence higher age resolution. 

For LAMOST, two peaks are visible in the underlying age distributions for metallicities $\feh \ge -0.2$\,dex: one at around $\tau \approx 9.35$ and one at $\tau \approx 9.83$. The latter peak becomes dominant and shifts to higher $\tau$ values, as $\feh$ increases. Similarly, for APOGEE a secondary peak is visible in the underlying age distributions for metallicities $\feh \ge -0.2$\,dex at around $\tau \approx 8.9$. 
In both cases the result is that the mean age for bins with $\feh \ge +0.2$\,dex is higher than that for solar-like metallicities.
This does not seem to be physical -- we expect more metal rich stars to be systematically younger. 
This might be attributed to the limitations of the method listed above in the \autoref{sec:full_survey}. An alternative explanation is that the effect is physical, and is caused by the fact that metal rich ($\feh \ge +0.2$\,dex) originate in the inner part of the Galaxy, and it takes a certain amount of time for the migration process to bring them to the solar vicinity, where they can be observed \citep[see][]{2018MNRAS.481.1645M}. Thus a small deficit of young metal rich stars can be expected.

It is very likely that all three effects are acting simultaneously. For better results, selection effects (like those presented in \citet{Selection}) have to be taken into account. Binning the data in distance or galactic latitude is also desirable. Both tasks are beyond the scope of the current work. 

Overall, the underlying age distributions obtained through the inversion allows to trace age-metallicity trends that are much less obvious if log(age) PDFs are considered. It also allows to remove structures like low-age spikes, that do not seem to be real and are likely due to observational scatter in physical parameters of older stars.

\subsection{Metallicity-age and age-metallicity relations}
Results from \autoref{sec:monometallicity} allow us to study metallicity-age and age-metallicity relations for RAVE-on, LAMOST and APOGEE.
Both relations can be defined through a two-dimensional distribution function $C(\tau, \feh) = C_{\feh}(\tau) N(\feh)$, where $C_{\feh}(\tau)$ is the log(age) PDF in the bin with metallicity $\feh$, and $N(\feh)$ is the number of stars in the survey as a function of metallicity. From that we can define log(age) as a function of metallicity as:
\begin{equation}
\tau(\feh) = \frac{\int_{\tau_{min}}^{\tau_{max}} C(\tau, \feh) d\,\tau}{\tau_{max} - \tau_{min}} , \label{eq:tau_feh}
\end{equation}
and similarly metallicity as a function of log(age):
\begin{equation}
\feh(\tau) = \frac{\int_{\feh_{min}}^{\feh_{max}} C(\tau, \feh) d\,\feh}{\feh_{max} - \feh_{min}}.\label{eq:feh_tau}
\end{equation}
Here, $(\tau_{min}, \tau_{max})$ is the considered range in log(age) and $(\feh_{min}, \feh_{max})$ is the range in metallicity.

\begin{figure}
	\myimage{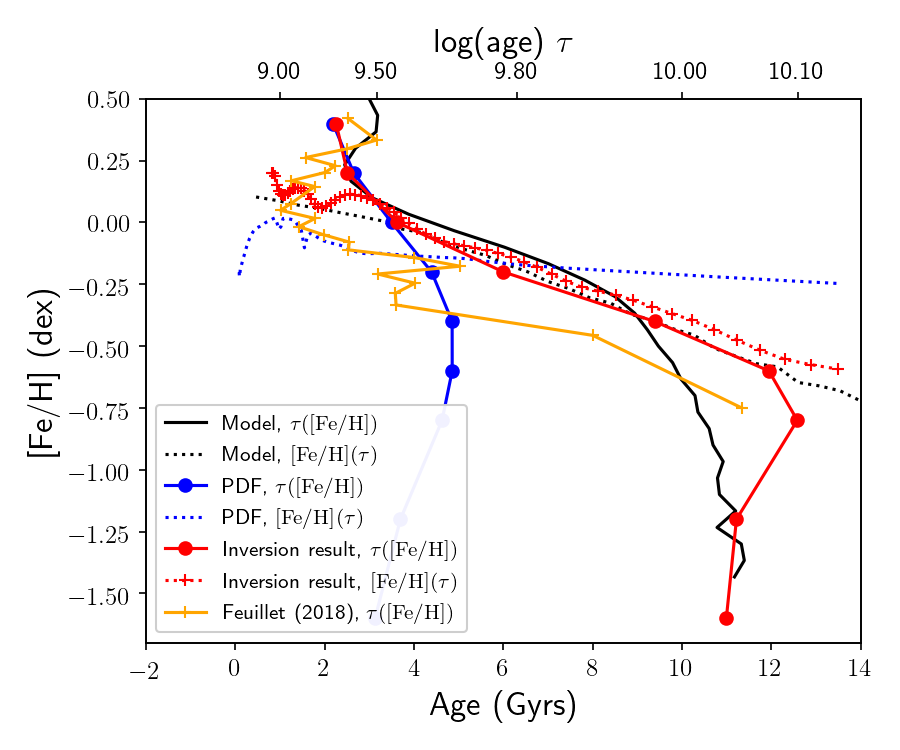}
	\caption{Age-metallicity ($\tau(\feh)$) relations for nearby APOGEE stars \citep[orange line,][]{2018MNRAS.477.2326F} and for RAVE-on stars (solid lines, this work). Blue lines are values derived from log(age) PDFs, red lines -- from the underlying age distributions obtained from inversion, black lines --  \citet{2013A&A...558A...9M} chemical evolution models. Dotted lines show mean metallicity as a function of log(age) for the same data.}  \label{fig:feh_age_rave_on}
\end{figure}

\begin{figure}
	\myimage{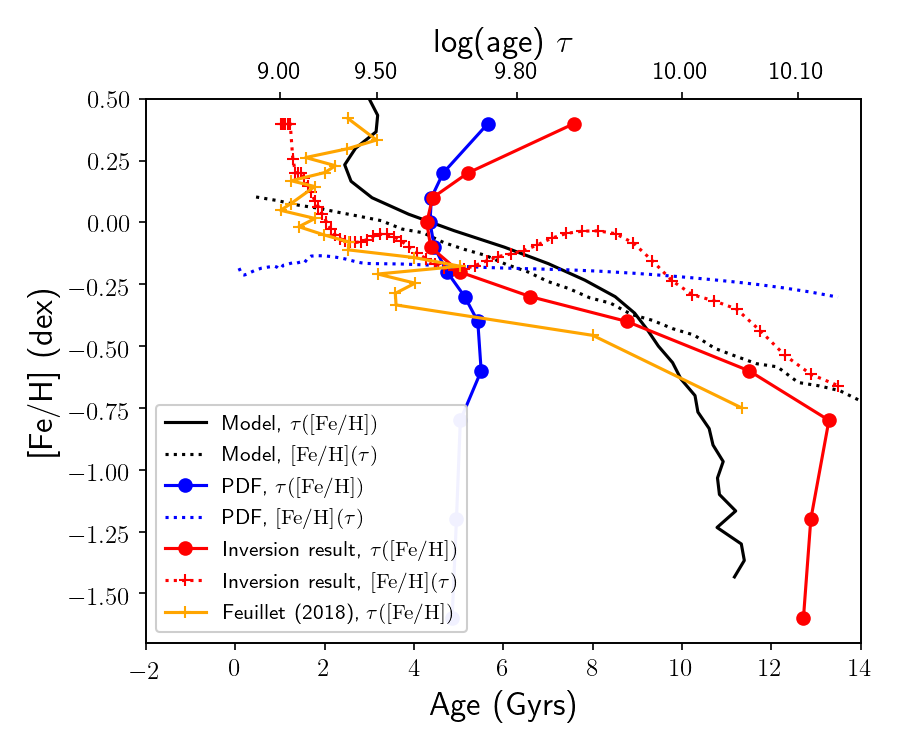}
	\caption{Same as \autoref{fig:feh_age_rave_on}, now for LAMOST stars} \label{fig:feh_age_lamost}
\end{figure}

\begin{figure}
	\myimage{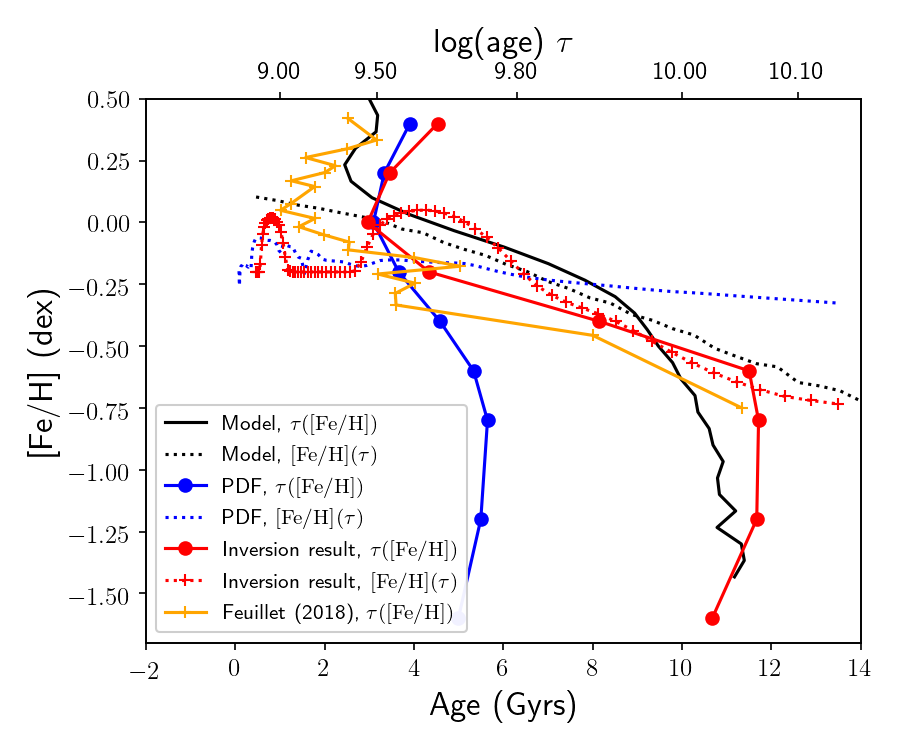}
	\caption{Same as \autoref{fig:feh_age_rave_on}, now for APOGEE stars} \label{fig:feh_age_apogee}
\end{figure}

In Figs. \ref{fig:feh_age_rave_on}, \ref{fig:feh_age_lamost} and \ref{fig:feh_age_apogee} we show $\tau(\feh)$ with blue solid and $\feh(\tau)$ with blue dotted lines. In these plots we will continue to use solid lines for $\tau(\feh)$ relations and dotted lines for $\feh(\tau)$ relations.
These two lines are nearly orthogonal, which is the consequence of the $C(\tau, \feh)$ distribution being very broad. 

We can now replace log(age) PDFs $C(\tau, \feh)$ with the underlying age distribution from the inversion $N(\tau, \feh)$ in \autoref{eq:tau_feh} and \ref{eq:feh_tau}. The results are shown in Figs. \ref{fig:feh_age_rave_on}, \ref{fig:feh_age_lamost} and \ref{fig:feh_age_apogee} with red solid and dotted lines. They are a lot closer to each other, which is a sign of a much tighter $C(\tau, \feh)$ function.

These data were compared to the results of the recent work by \citet{2018MNRAS.477.2326F}, who presented the analysis of a sample of 721 nearby red giant stars selected from APOGEE \citep{APOGEE}. All these stars are closer than 400 pc and have reliable TGAS parallaxes, which allows to determine log(age) to 0.07\,dex precision. This sample was used, among other applications, to derive mean ages as a function of metallicity, shown in Figs. \ref{fig:feh_age_rave_on}, \ref{fig:feh_age_lamost} and \ref{fig:feh_age_apogee} with orange solid line. 
We note that the data presented in \citet{2018MNRAS.477.2326F}, as well as our results for $\tau(\feh)$ give age distribution and mean age in each metallicity bin, thus measuring age as a function of metallicity $\tau(\feh)$. At the same time, models of chemical evolution typically focus on the metallicity as a function of age $\feh(\tau)$ \citep[see for example Fig. 4 in ][]{2013A&A...558A...9M}, which is strictly speaking a different function. Functions $\tau(\feh)$ and $\feh(\tau)$ can be close to each other only if the two-dimensional distribution $C(\tau, \feh)$ is tight. Hence in a general case care must be taken in comparing  $\tau(\feh)$ and $\feh(\tau)$, as it might lead to wrong conclusions.

If we compare the age-metallicity distributions from log(age) PDFs $C(\tau, \feh)$ with those obtained from the underlying age distributions computed through the inversion, $N(\tau, \feh)$, we see that the latter are much closer to \citet{2018MNRAS.477.2326F}, than the former. The small systematic offset can be attributed to the fact that, as opposed to the solar neighbourhood APOGEE subsample used by \citet{2018MNRAS.477.2326F}, surveys used here contain more thick-disk and halo stars that are systematically older.

We can compare our age-metallicity and metallicity-age trends with those predicted by chemical evolution models. We use the model described in \citet{2013A&A...558A...9M}. We added an additional smoothing to the two-dimensional age-metallicity distribution predicted by this model. The smoothing scale was chosen to be close to the typical uncertainty in age (2 Gyrs) and metallicity (0.1 dex). We then calculated $\tau(\feh)$ and $\feh(\tau)$ for this distribution and show it in Figs. \ref{fig:feh_age_rave_on}, \ref{fig:feh_age_lamost} with black solid and \ref{fig:feh_age_apogee} with black dotted lines. Note that mean age $\tau$ starts to increase as a function of $\feh$ for $\feh > 0$ for the model data as well as LAMOST and APOGEE inversion results. This is very likely caused by the absence of young metal-rich stars in the solar vicinity, which are formed in the inner Galaxy and need time to migrate outwards. Mean ages for metal poor stars are also systematically larger for inversion results than those predicted by models. This might be caused by the fact that UniDAM allows for stellar ages up to the age of the Universe ($\approx 13.5\,$Gyrs), while the maximum stellar age in the \citet{2013A&A...558A...9M} model is $11.175\,$Gyrs (before the smoothing was applied). In general, on all three plots $\tau(\feh)$ and $\feh(\tau)$ from the model and from age inversion results are close to each other.


\section{Summary and outlook}
In this work we present age inversions -- a method of revealing the underlying distribution of a large ($N > 10^3$) ensembles of stars in log(age) from their cumulative log(age) PDF produced by UniDAM. This allows us to remove biases inherent to isochrone fitting. The method was tested on simulated data showing that it allows us to reconstruct the underlying age distribution of stars in the simulated sample. The inversion results are a lot closer to the real underlying age distribution than the stacked log(age) PDF or the distribution of mean PDF values.

The method was further applied to data produced by UniDAM for real surveys, deriving different age distributions for different surveys. This is expected, as surveys use different observational strategies and focus at different samples of stars. Systematic offsets between survey data and PARSEC models used in this work and survey pipeline artefacts are the main limitations of the method. Due to this we are limited in our ability to generate mono-age populations that will be distributed closely to the survey stars, which is needed for a reliable age inversion. 

We also apply age inversion method to mono-metallicity slices of RAVE-on, LAMOST and APOGEE surveys. This allows us to trace how the age distribution changes as a function of metallicity, and to reconstruct both metallicity-age and age-metallicity relations for RAVE-on, LAMOST  and APOGEE samples, successfully removing artefacts of the isochrone fitting. We obtain results similar to those published in \citet{2018MNRAS.477.2326F} for a much smaller high-precision APOGEE sample and to those predicted by chemical evolution models.
The number of stars in both RAVE-on and LAMOST surveys, at least for metallicities close to solar, is high enough to make further division in, for example, galactic latitude, distance or $\alpha$-abundance bins. This is considered beyond the scope of the current work.

In the future, systematic offsets between data and models are expected to decrease as the uncertainties (both random and systematic) of observations decrease and as models improve. Furthermore, the future spectroscopic surveys such as 4MOST \citep{FourMOST} and WEAVE \citep{WEAVE} will provide data for millions of stars. The age inversion method presented here will be suited to provide the underlying age distribution of such large survey data sets as a whole as well as as functions of, for example, metallicity, location in the Galaxy and $\alpha$-elements abundances. 

One can be tempted to use the vast amount of data contained in Gaia to derive age distributions from photometry alone, like it was done, for example, in \citet{Dolphin_2013}. However, it might be difficult given the lack of metallicity information in photometric data, which is essential for unambiguous age determination. In any case, such kind of study will differ substantially from that presented in this work, and the new method to solve \autoref{eq:diff} might be needed.

It is possible to make use of Gaia parallax data to improve the precision of ages derived from spectrophotometric data, as it was done in \citet{UniDAM2} and \citet{UniDAM2a}. However, this will require a more complex simulation strategy that will take into account proper parallax uncertainty distribution. Apart from that, Gaia parallax zero-point offset \citep[see, for example, ][]{2019arXiv190208634L} have to be treated properly, as it can inflict a systematic age bias.

\section*{Acknowledgements}
Authors thank the SAGE group at Max Planck Institute for Solar System Research for many fruitful discussions.
Authors thank the anonymous referee for a detailed report with many useful suggestions, which helped us to improve the manuscript substantially.

The research leading to the presented results has received funding from the European Research Council under the European Community’s Seventh Framework Programme (FP7/2007-2013)/ERC grant agreement (No 338251, StellarAges).

This research made use of Astropy, a community-developed core Python package for Astronomy \citep{2013A&A...558A..33A}.
This research made use of matplotlib, a Python library for publication quality graphics \citep{Hunter:2007}.
This research made use of SciPy \citep{jones_scipy_2001}.
This research made use of TOPCAT, an interactive graphical viewer and editor for tabular data \citep{2005ASPC..347...29T}.
Funding for RAVE has been provided by: the Australian Astronomical Observatory; the Leibniz-Institut fuer Astrophysik Potsdam (AIP); the Australian National University; the Australian Research Council; the French National Research Agency; the German Research Foundation (SPP 1177 and SFB 881); the European Research Council (ERC-StG 240271 Galactica); the Istituto Nazionale di Astrofisica at Padova; The Johns Hopkins University; the National Science Foundation of the USA (AST-0908326); the W. M. Keck foundation; the Macquarie University; the Netherlands Research School for Astronomy; the Natural Sciences and Engineering Research Council of Canada; the Slovenian Research Agency; the Swiss National Science Foundation; the Science \& Technology Facilities Council of the UK; Opticon; Strasbourg Observatory; and the Universities of Groningen, Heidelberg and Sydney. The RAVE web site is at \url{https://www.rave-survey.org}.
Funding for the Sloan Digital Sky Survey IV has been provided by the Alfred P. Sloan Foundation, the U.S. Department of Energy Office of Science, and the Participating Institutions. SDSS-IV acknowledges support and resources from the Center for High-Performance Computing at the University of Utah. The SDSS web site is www.sdss.org. SDSS-IV is managed by the Astrophysical Research Consortium for the Participating Institutions of the SDSS Collaboration including the Brazilian Participation Group, the Carnegie Institution for Science, Carnegie Mellon University, the Chilean Participation Group, the French Participation Group, Harvard-Smithsonian Center for Astrophysics, Instituto de Astrof\'isica de Canarias, The Johns Hopkins University, Kavli Institute for the Physics and Mathematics of the Universe (IPMU) / University of Tokyo, Lawrence Berkeley National Laboratory, Leibniz Institut f\"ur Astrophysik Potsdam (AIP), Max-Planck-Institut f\"ur Astronomie (MPIA Heidelberg), Max-Planck-Institut f\"ur Astrophysik (MPA Garching), Max-Planck-Institut f\"ur Extraterrestrische Physik (MPE), National Astronomical Observatories of China, New Mexico State University, New York University, University of Notre Dame, Observat\'ario Nacional / MCTI, The Ohio State University, Pennsylvania State University, Shanghai Astronomical Observatory, United Kingdom Participation Group, Universidad Nacional Aut\'onoma de M\'exico, University of Arizona, University of Colorado Boulder, University of Oxford, University of Portsmouth, University of Utah, University of Virginia, University of Washington, University of Wisconsin, Vanderbilt University, and Yale University.
Guoshoujing Telescope (the Large Sky Area Multi-Object Fiber Spectroscopic Telescope LAMOST) is a National Major Scientific Project built by the Chinese Academy of Sciences. Funding for the project has been provided by the National Development and Reform Commission. LAMOST is operated and managed by the National Astronomical Observatories, Chinese Academy of Sciences.
The GALAH survey is based on observations made at the Australian Astronomical Observatory, under programmes A/2013B/13, A/2014A/25, A/2015A/19, A/2017A/18. We acknowledge the traditional owners of the land on which the AAT stands, the Gamilaraay people, and pay our respects to elders past and present. 
\bibliographystyle{aa.bst}
\bibliography{sage_age_inversion.bib}

\begin{thebibliography}{31}
\expandafter\ifx\csname natexlab\endcsname\relax\def\natexlab#1{#1}\fi

\bibitem[{{Astropy Collaboration} {et~al.}(2013){Astropy Collaboration},
  {Robitaille}, {Tollerud}, {Greenfield}, {Droettboom}, {Bray}, {Aldcroft},
  {Davis}, {Ginsburg}, {Price-Whelan}, {Kerzendorf}, {Conley}, {Crighton},
  {Barbary}, {Muna}, {Ferguson}, {Grollier}, {Parikh}, {Nair}, {Unther},
  {Deil}, {Woillez}, {Conseil}, {Kramer}, {Turner}, {Singer}, {Fox}, {Weaver},
  {Zabalza}, {Edwards}, {Azalee Bostroem}, {Burke}, {Casey}, {Crawford},
  {Dencheva}, {Ely}, {Jenness}, {Labrie}, {Lim}, {Pierfederici}, {Pontzen},
  {Ptak}, {Refsdal}, {Servillat}, \& {Streicher}}]{2013A&A...558A..33A}
{Astropy Collaboration}, {Robitaille}, T.~P., {Tollerud}, E.~J., {et~al.} 2013,
  \aap, 558, A33

\bibitem[{{Bressan} {et~al.}(2012){Bressan}, {Marigo}, {Girardi}, {Salasnich},
  {Dal Cero}, {Rubele}, \& {Nanni}}]{PARSEC}
{Bressan}, A., {Marigo}, P., {Girardi}, L., {et~al.} 2012, \mnras, 427,
  127–145

\bibitem[{{Casey} {et~al.}(2016){Casey}, {Hawkins}, {Hogg}, {Ness},
  {Walter-Rix}, {Kordopatis}, {Kunder}, {Steinmetz}, {Koposov}, {Enke},
  {Sanders}, {Gilmore}, {Zwitter}, {Freeman}, {Casagrande}, {Matijevič},
  {Seabroke}, {Bienaymé}, {Bland-Hawthorn}, {Gibson}, {Grebel}, {Helmi},
  {Munari}, {Navarro}, {Reid}, {Siebert}, \& {Wyse}}]{2016arXiv160902914C}
{Casey}, A.~R., {Hawkins}, K., {Hogg}, D.~W., {et~al.} 2016, ArXiv e-prints
  [\eprint[arXiv]{1609.02914}]

\bibitem[{{Cutri, R.~M. et al.}(2014)}]{2014yCat.2328....0C}
{Cutri, R.~M. et al.} 2014, VizieR Online Data Catalog, 2328

\bibitem[{{Dalton} {et~al.}(2014){Dalton}, {Trager}, {Abrams}, {Bonifacio},
  {L{\'o}pez Aguerri}, {Middleton}, {Benn}, {Dee}, {Say{\`e}de}, {Lewis},
  {Pragt}, {Pico}, {Walton}, {Rey}, {Allende Prieto}, {Pe{\~n}ate}, {Lhome},
  {Ag{\'o}cs}, {Alonso}, {Terrett}, {Brock}, {Gilbert}, {Ridings}, {Guinouard},
  {Verheijen}, {Tosh}, {Rogers}, {Steele}, {Stuik}, {Tromp}, {Jasko}, {Kragt},
  {Lesman}, {Mottram}, {Bates}, {Gribbin}, {Fernand o Rodriguez}, {Delgado},
  {Martin}, {Cano}, {Navarro}, {Irwin}, {Lewis}, {Gonzalez Solares},
  {O'Mahony}, {Bianco}, {Zurita}, {ter Horst}, {Molinari}, {Lodi}, {Guerra},
  {Vallenari}, \& {Baruffolo}}]{WEAVE}
{Dalton}, G., {Trager}, S., {Abrams}, D.~C., {et~al.} 2014, in Society of
  Photo-Optical Instrumentation Engineers (SPIE) Conference Series, Vol. 9147,
  Ground-based and Airborne Instrumentation for Astronomy V, 91470L

\bibitem[{{de Jong} {et~al.}(2016){de Jong}, {Barden}, {Bellido-Tirado},
  {Brynnel}, {Frey}, {Giannone}, {Haynes}, {Johl}, {Phillips}, {Schnurr},
  {Walcher}, {Winkler}, {Ansorge}, {Feltzing}, {McMahon}, {Baker}, {Caillier},
  {Dwelly}, {Gaessler}, {Iwert}, {Mandel}, {Piskunov}, {Pragt}, {Walton},
  {Bensby}, {Bergemann}, {Chiappini}, {Christlieb}, {Cioni}, {Driver},
  {Finoguenov}, {Helmi}, {Irwin}, {Kitaura}, {Kneib}, {Liske}, {Merloni},
  {Minchev}, {Richard}, \& {Starkenburg}}]{FourMOST}
{de Jong}, R.~S., {Barden}, S.~C., {Bellido-Tirado}, O., {et~al.} 2016, in
  Society of Photo-Optical Instrumentation Engineers (SPIE) Conference Series,
  Vol. 9908, Ground-based and Airborne Instrumentation for Astronomy VI, 99081O

\bibitem[{Dolphin(2013)}]{Dolphin_2013}
Dolphin, A.~E. 2013, The Astrophysical Journal, 775, 76

\bibitem[{{Feuillet} {et~al.}(2016){Feuillet}, {Bovy}, {Holtzman}, {Girardi},
  {MacDonald}, {Majewski}, \& {Nidever}}]{2016ApJ...817...40F}
{Feuillet}, D.~K., {Bovy}, J., {Holtzman}, J., {et~al.} 2016, \apj, 817, 40

\bibitem[{{Feuillet} {et~al.}(2018){Feuillet}, {Bovy}, {Holtzman}, {Weinberg},
  {García-Hernández}, {Hearty}, {Majewski}, {Roman-Lopes}, {Rybizki}, \&
  {Zamora}}]{2018MNRAS.477.2326F}
{Feuillet}, D.~K., {Bovy}, J., {Holtzman}, J., {et~al.} 2018, \mnras, 477,
  2326–2348

\bibitem[{Hunter(2007)}]{Hunter:2007}
Hunter, J.~D. 2007, Computing In Science \& Engineering, 9, 90

\bibitem[{Lawson \& Hanson(1995)}]{NNLS}
Lawson, C.~L. \& Hanson, R.~J. 1995, {Solving least squares problems}, [rev.
  ed.] edn. (Philadelphia : SIAM)

\bibitem[{{Leung} \& {Bovy}(2019)}]{2019arXiv190208634L}
{Leung}, H.~W. \& {Bovy}, J. 2019, arXiv e-prints, arXiv:1902.08634

\bibitem[{{Luo} {et~al.}(2015){Luo}, {Zhao}, {Zhao}, {Deng}, {Liu}, {Jing},
  {Wang}, {Zhang}, {Shi}, {Cui}, {Chu}, {Li}, {Bai}, {Wu}, {Cai}, {Cao}, {Cao},
  {Carlin}, {Chen}, {Chen}, {Chen}, {Chen}, {Chen}, {Chen}, {Chen},
  {Christlieb}, {Chu}, {Cui}, {Dong}, {Du}, {Fan}, {Feng}, {Fu}, {Gao}, {Gong},
  {Gu}, {Guo}, {Han}, {He}, {Hou}, {Hou}, {Hou}, {Hu}, {Hu}, {Hu}, {Huo},
  {Jia}, {Jiang}, {Jiang}, {Jiang}, {Jin}, {Kong}, {Kong}, {Lei}, {Li}, {Li},
  {Li}, {Li}, {Li}, {Li}, {Li}, {Li}, {Li}, {Li}, {Li}, {Li}, {Liang}, {Lin},
  {Liu}, {Liu}, {Liu}, {Liu}, {Lu}, {Luo}, {Mao}, {Newberg}, {Ni}, {Qi}, {Qi},
  {Shen}, {Shi}, {Song}, {Song}, {Su}, {Su}, {Tang}, {Tao}, {Tian}, {Wang},
  {Wang}, {Wang}, {Wang}, {Wang}, {Wang}, {Wang}, {Wang}, {Wang}, {Wang},
  {Wang}, {Wang}, {Wang}, {Wang}, {Wang}, {Wang}, {Wang}, {Wang}, {Wang},
  {Wang}, {Wei}, {Wei}, {Wu}, {Wu}, {Wu}, {Wu}, {Xing}, {Xu}, {Xu}, {Xu},
  {Yan}, {Yang}, {Yang}, {Yang}, {Yang}, {Yao}, {Yu}, {Yuan}, {Yuan}, {Yuan},
  {Yuan}, {Zhai}, {Zhang}, {Zhang}, {Zhang}, {Zhang}, {Zhang}, {Zhang},
  {Zhang}, {Zhang}, {Zhao}, {Zhou}, {Zhou}, {Zhu}, {Zhu}, {Zou}, \&
  {Zuo}}]{LAMOST}
{Luo}, A.-L., {Zhao}, Y.-H., {Zhao}, G., {et~al.} 2015, Research in Astronomy
  and Astrophysics, 15, 1095

\bibitem[{{Majewski} {et~al.}(2017){Majewski}, {Schiavon}, {Frinchaboy},
  {Allende Prieto}, {Barkhouser}, {Bizyaev}, {Blank}, {Brunner}, {Burton},
  {Carrera}, {Chojnowski}, {Cunha}, {Epstein}, {Fitzgerald}, {García Pérez},
  {Hearty}, {Henderson}, {Holtzman}, {Johnson}, {Lam}, {Lawler}, {Maseman},
  {Mészáros}, {Nelson}, {Nguyen}, {Nidever}, {Pinsonneault}, {Shetrone},
  {Smee}, {Smith}, {Stolberg}, {Skrutskie}, {Walker}, {Wilson}, {Zasowski},
  {Anders}, {Basu}, {Beland}, {Blanton}, {Bovy}, {Brownstein}, {Carlberg},
  {Chaplin}, {Chiappini}, {Eisenstein}, {Elsworth}, {Feuillet}, {Fleming},
  {Galbraith-Frew}, {García}, {García-Hernández}, {Gillespie}, {Girardi},
  {Gunn}, {Hasselquist}, {Hayden}, {Hekker}, {Ivans}, {Kinemuchi}, {Klaene},
  {Mahadevan}, {Mathur}, {Mosser}, {Muna}, {Munn}, {Nichol}, {O'Connell},
  {Parejko}, {Robin}, {Rocha-Pinto}, {Schultheis}, {Serenelli}, {Shane}, {Silva
  Aguirre}, {Sobeck}, {Thompson}, {Troup}, {Weinberg}, \& {Zamora}}]{APOGEE}
{Majewski}, S.~R., {Schiavon}, R.~P., {Frinchaboy}, P.~M., {et~al.} 2017, \aj,
  154, 94

\bibitem[{{Martell} {et~al.}(2016){Martell}, {Sharma}, {Buder}, {Duong},
  {Schlesinger}, {Simpson}, {Lind}, {Ness}, {Marshall}, {Asplund},
  {Bland-Hawthorn}, {Casey}, {De Silva}, {Freeman}, {Kos}, {Lin}, {Zucker},
  {Zwitter}, {Anguiano}, {Bacigalupo}, {Carollo}, {Casagrande}, {Da Costa},
  {Horner}, {Huber}, {Hyde}, {Kaﬂe}, {Lewis}, {Nataf}, {Stello}, {Tinney},
  {Watson}, \& {Wittenmyer}}]{GALAH}
{Martell}, S., {Sharma}, S., {Buder}, S., {et~al.} 2016, ArXiv e-prints
  [\eprint[arXiv]{1609.02822}]

\bibitem[{{Minchev} {et~al.}(2018){Minchev}, {Anders}, {Recio-Blanco},
  {Chiappini}, {de Laverny}, {Queiroz}, {Steinmetz}, {Adibekyan}, {Carrillo},
  \& {Cescutti}}]{2018MNRAS.481.1645M}
{Minchev}, I., {Anders}, F., {Recio-Blanco}, A., {et~al.} 2018, \mnras, 481,
  1645

\bibitem[{{Minchev} {et~al.}(2013){Minchev}, {Chiappini}, \&
  {Martig}}]{2013A&A...558A...9M}
{Minchev}, I., {Chiappini}, C., \& {Martig}, M. 2013, \aap, 558, A9

\bibitem[{{Minchev} {et~al.}(2019){Minchev}, {Matijevic}, {Hogg}, {Guiglion},
  {Steinmetz}, {Anders}, {Chiappini}, {Martig}, {Queiroz}, \&
  {Scannapieco}}]{2019MNRAS.487.3946M}
{Minchev}, I., {Matijevic}, G., {Hogg}, D.~W., {et~al.} 2019, \mnras, 487, 3946

\bibitem[{{Mints}(2018)}]{UniDAM2a}
{Mints}, A. 2018, arXiv e-prints, arXiv:1805.01640

\bibitem[{{Mints} \& {Hekker}(2017)}]{UniDAM1}
{Mints}, A. \& {Hekker}, S. 2017, \aap, 604, A108

\bibitem[{{Mints} \& {Hekker}(2018)}]{UniDAM2}
{Mints}, A. \& {Hekker}, S. 2018, \aap, 618, A54

\bibitem[{{Mints} \& {Hekker}(2019)}]{Selection}
{Mints}, A. \& {Hekker}, S. 2019, \aap, 621, A17

\bibitem[{{Perryman} {et~al.}(2001){Perryman}, {de Boer}, {Gilmore}, {Høg},
  {Lattanzi}, {Lindegren}, {Luri}, {Mignard}, {Pace}, \& {de
  Zeeuw}}]{2001A&A...369..339P}
{Perryman}, M.~A.~C., {de Boer}, K.~S., {Gilmore}, G., {et~al.} 2001, \aap,
  369, 339–363

\bibitem[{{Queiroz} {et~al.}(2018){Queiroz}, {Anders}, {Santiago}, {Chiappini},
  {Steinmetz}, {Ponte}, {Stassun}, {da Costa}, {Maia}, {Crestani}, {Beers},
  {Fernández-Trincado}, {García-Hernández}, {Roman-Lopes}, \&
  {Zamora}}]{2018MNRAS.tmp..326Q}
{Queiroz}, A.~B.~A., {Anders}, F., {Santiago}, B.~X., {et~al.} 2018, \mnras
  [\eprint[arXiv]{1710.09970}]

\bibitem[{{Scipy team}(2001)}]{jones_scipy_2001}
{Scipy team}. 2001, {SciPy}: Open source scientific tools for Python

\bibitem[{{Skrutskie} {et~al.}(2006){Skrutskie}, {Cutri}, {Stiening},
  {Weinberg}, {Schneider}, {Carpenter}, {Beichman}, {Capps}, {Chester},
  {Elias}, {Huchra}, {Liebert}, {Lonsdale}, {Monet}, {Price}, {Seitzer},
  {Jarrett}, {Kirkpatrick}, {Gizis}, {Howard}, {Evans}, {Fowler}, {Fullmer},
  {Hurt}, {Light}, {Kopan}, {Marsh}, {McCallon}, {Tam}, {Van Dyk}, \&
  {Wheelock}}]{2006AJ....131.1163S}
{Skrutskie}, M.~F., {Cutri}, R.~M., {Stiening}, R., {et~al.} 2006, \aj, 131,
  1163–1183

\bibitem[{{Soderblom}(2010)}]{2010ARA&A..48..581S}
{Soderblom}, D.~R. 2010, \araa, 48, 581–629

\bibitem[{{Taylor}(2005)}]{2005ASPC..347...29T}
{Taylor}, M.~B. 2005, in Astronomical Society of the Pacific Conference Series,
  Vol. 347, Astronomical Data Analysis Software and Systems XIV, ed.
  P.~{Shopbell}, M.~{Britton}, \& R.~{Ebert}, 29

\bibitem[{{Tucci Maia} {et~al.}(2016){Tucci Maia}, {Ram{\'\i}rez},
  {Mel{\'e}ndez}, {Bedell}, {Bean}, \& {Asplund}}]{2016A&A...590A..32T}
{Tucci Maia}, M., {Ram{\'\i}rez}, I., {Mel{\'e}ndez}, J., {et~al.} 2016, \aap,
  590, A32

\bibitem[{{Wu} {et~al.}(2017){Wu}, {Xiang}, {Zhang}, {Li}, {Bi}, {Liu}, {Fu},
  {Huang}, {Tian}, \& {Liu}}]{2017RAA....17....5W}
{Wu}, Y.-Q., {Xiang}, M.-S., {Zhang}, X.-F., {et~al.} 2017, Research in
  Astronomy and Astrophysics, 17, 5

\bibitem[{Xiang {et~al.}(2017)Xiang, Liu, Shi, Yuan, Huang, Chen, Wang, Tian,
  Wu, Yang, Zhang, Huo, \& Ren}]{Xiang_2017}
Xiang, M., Liu, X., Shi, J., {et~al.} 2017, The Astrophysical Journal
  Supplement Series, 232, 2

\end{thebibliography}

\end{document}